\documentclass[12pt, draftclsnofoot, onecolumn]{IEEEtran}
\pdfoutput=1
\usepackage[noadjust]{cite}
\usepackage{amsmath}
\usepackage{array}
\usepackage{mdwmath}
\usepackage{amssymb}
\usepackage{algorithm}
\usepackage[noend]{algpseudocode}
\usepackage{eqparbox}
\usepackage{stfloats}
\usepackage{url}
\usepackage[pdftex]{graphicx}
\graphicspath{{../pdf/}{../jpeg/}}
\DeclareGraphicsExtensions{.pdf,.jpeg,.png}
\usepackage{epstopdf}
\usepackage{amsfonts}

\usepackage{color}

\usepackage{tikz}
\usepackage{pgfplots}
\usepackage{subcaption}

\newtheorem{lemma}{Lemma}
\newtheorem{theorem}{Theorem}
\newtheorem{remark}{Remark}
\newtheorem{corollary}{Corollary}

\begin{document}

\title{Multi-User Opportunistic Beamforming using Reconfigurable Surfaces}
\author{Qurrat-Ul-Ain~Nadeem,~\IEEEmembership{Member,~IEEE,} Anas~Chaaban,~\IEEEmembership{Senior Member,~IEEE,}
        M{\'e}rouane~Debbah,~\IEEEmembership{Fellow,~IEEE,} 
\thanks{Q.-U.-A. Nadeem and A. Chaaban are with School of Engineering, The University of British Columbia, Kelowna, Canada.  (e-mail:  \{qurrat.nadeem, anas.chaaban\}@ubc.ca)}
\thanks{M. Debbah is  with  CentraleSup{\'e}lec, Gif-sur-Yvette, France and Mathematical and Algorithmic Sciences Lab, Huawei France R\&D, Paris, France (e-mail: merouane.debbah@huawei.com, merouane.debbah@centralesupelec.fr).}
}

\markboth{}%
{Shell \MakeLowercase{\textit{et al.}}: Bare Demo of IEEEtran.cls for Journals}
\maketitle
\vspace{-.6in}
\begin{abstract}

Multi-user (MU) diversity yields sum-rate gains by scheduling a user for transmission at times when its channel is near its peak. The only information required at the base station (BS) for scheduling is the users' signal-to-noise ratios (SNR)s. MU diversity gains are limited in environments with line-of-sight (LoS) channel components and/or spatial correlation. To remedy this, previous works have proposed opportunistic beamforming (OBF) using multiple antennas at the BS to transmit the same signal, modulated by time-varying gains, to the best user at each time slot. In this paper, we propose reconfigurable surface (RS)-assisted OBF to increase the range of channel fluctuations in a single-antenna broadcast channel (BC), where opportunistic scheduling (OS) strategy achieves the sum-rate capacity. The RS is abstracted as an array of passive reflecting elements, and is dumb in the sense that it only induces random phase shifts onto the impinging electromagnetic waves, without requiring any channel state information. We develop the sum-rate scaling laws  under Rayleigh, Rician and correlated Rayleigh fading and show that RS-assisted OBF with only a single-antenna BS outperforms multi-antenna BS-assisted OBF. We also extend our results to OFDMA systems and the multi-antenna BC.

\end{abstract}

\vspace{-.2in}
\section{Introduction}

Multi-path fading is generally viewed as a source of unreliability and is mitigated using diversity techniques over time, frequency or space. In addition to these classical diversity modes, foundational works in information theory \cite{MUdiv, MUdiv1, dumb} have shown that in a wireless system with multiple active users sharing the transmission medium, multi-user (MU) diversity can be realized using a scheduler at the medium-access control (MAC) layer, which schedules the user having the best channel quality during a coherence interval. This requires instantaneous channel state information at the transmitter (CSIT) in the form of the signal-to-noise ratio (SNR) of all users. MU-diversity gains then arise from the fact that in a system with many users whose channels fade independently, there is likely to be a user at each time whose channel is near its peak.

The concept of MU diversity is best motivated by an information-theoretic result by Knopp and Humblet in \cite{MUdiv} focusing on an uplink scenario with multiple users communicating with a single-antenna base station (BS) via fading channels. Their work shows that the {\it average sum-rate capacity} of this system, defined as the maximum achievable sum of the long-term average data rates of all users, is achieved by a simple opportunistic scheduling (OS) scheme which schedules at one time only the user with the largest SNR. Similar results have been obtained for the downlink from a single-antenna BS to mobile users in \cite{MUdiv1, dumb}, where the average sum-rate capacity (referred to henceforth as sum-capacity) is achieved by an OS scheme which transmits to the user with the largest SNR. As a consequence, MU-diversity gains arise and can be explained as: in a system with many users with independently time-varying channels, it is likely to have a user whose SNR is much stronger than the average SNR at each time. By scheduling this user, the independent fading of the users' channels is exploited and the spectral efficiency can be made significantly higher than that of a non-faded channel with the same average SNR. 


From a MU-diversity perspective, fading is considered as a source of randomization which can be exploited to yield significant gains. As the number of users increases, it becomes more likely to schedule users when they are near their channel peaks through OS. The larger the dynamic range of the channel fluctuations, the higher the peaks and the larger the MU-diversity gain. In practice, such gains are limited in two ways. First, the channel may fade slowly as compared to the delay tolerance of the application. Second, even under fast fading (relative to the latency time-scale), there may be dominant line-of-sight (LoS) paths or high saptial correlation in the channel. Both these factors reduce the dynamic range of channel fluctuations \cite{dumb, Ric, slow_f}.


To improve MU-diversity gains in these scenarios, Viswanath \textit{et al.} \cite{dumb} proposed an opportunistic beamforming (OBF) scheme, where multiple dumb antennas are used at the BS to transmit weighted replicas of the same signal, and the complex-valued weights are varied in a controlled but randomized way. This induces channel variations through constructive/destructive superposition of multi-paths. The SNR is tracked by each user and fed-back to the BS, which then schedules the best user for transmission in each time-slot. Large MU diversity gains can be realized using OBF, especially when the range of channel fluctuations is limited. It is important to note that even with multiple antennas at the BS, i.e., the multiple-input single-output (MISO) scenario, the authors retained OS strategy which is no longer capacity achieving.  However, as opposed to the optimal Dirty Paper Coding (DPC) which requires full CSIT, OBF only requires users' SNR feedback irrespective of the number of antennas.


Since the introduction of OBF, it has been applied in different contexts \cite{slow_f,lit1, lit2, lit3, lit4, lit5,Ric1, Ric, OBFOFDMA,OBFOFDMA1}, with several works focusing on designing fair schedulers \cite{Ric1, Ric}. The OBF scheme from \cite{dumb} can also be extended to wide-band channels that undergo frequency-selective fading, where several users can be scheduled on orthogonal frequency bands simultaneously using orthogonal frequency-division multiple access (OFDMA). By employing OBF and scheduling the strongest user at each frequency band, MU diversity  can be exploited in both the time and frequency domains. Several works have explored feedback reduction schemes for OBF in OFDMA systems~\cite{OBFOFDMA,OBFOFDMA1}. 

In this work, instead of relying on the use of multiple active antennas at the BS, we propose to deploy a dumb reconfigurable surface (RS) close to the BS to induce random fluctuations in the BS-user channels and enhance the MU-diversity gains. The RS is a planar array of nearly passive, low-cost, reflecting elements with reconfigurable parameters  \cite{Renzo2019, LIS, LIS1, LIS_los, my_IRS, my_IRS1,  spec1,RIS, RIS1,spec}. Each element independently introduces a phase shift onto the incoming electromagnetic waves, and desirable communication objectives can be achieved by smartly adjusting these phase shifts, while consuming very low energy due to the elements' passive nature. This renders this technology more energy efficient than existing methods for improving the user rates like cell-densification or deploying massive antenna arrays. Current implementations of RSs include reflect-arrays and reconfigurable metasurfaces \cite{RA_3, Renzo2019}.


In our proposed scheme, which we refer to as RS-assisted OBF, we keep a single-antenna at the BS and deploy a dumb RS consisting of $N$ passive reflecting elements in the LoS of the BS. The RS neither requires CSIT nor does it need to be `intelligent', since the applied phase shifts do not need to be optimized and are in fact selected randomly from the uniform distribution in most of the considered scenarios. Each user measures its downlink SNR based on a common pilot symbol transmitted by the BS, and feeds back this information to the BS. The BS then schedules the best user in each time-slot, which is sum-capacity optimal in a single-antenna broadcast channel (BC). We develop the sum-capacity scaling laws versus the number of users $K$ under RS-assisted OBF, when the RS-to-users channels undergo: 1) independent Rayleigh fading, 2) independent Rician fading, and 3) correlated Rayleigh fading. Then, we show that our scheme that uses only a single antenna at the BS achieves higher average sum-capacity scaling than that achieved by the BS-assisted OBF in \cite{dumb}. We also extend the derived results to wide-band channels where OFDMA is exploited to serve multiple users in each time-slot.

Finally, we extend the results to the MISO BC scenario under OS strategy, which is no longer capacity achieving as optimally multiple beams can be sent in the same time-slot to serve multiple users using beamforming. In fact, DPC combined with beamforming achieves the capacity of the multi-antenna BC, but it is computationally expensive and requires perfect CSIT for all users with respect to all the transmit (Tx) antennas. OS, on the other hand, requires just one SNR feedback per user irrespective of the number of antennas and is very simple to implement.  Therefore, like \cite{dumb,Ric1,Ric, slow_f, OBFOFDMA,lit4}, we retain the use of OS and study the impact of RS-assisted OBF on the sum-rate scaling of the MISO BC. This is the first work to study RS-assisted systems from an information-theoretic perspective and even though we focus on OS while providing some extensions for OFDMA systems, our results will be fundamental in the future in extending the analysis to space division multiple access (SDMA) systems, where multiple users are served in each time-slot using beamforming schemes such as zero-forcing and random beamforming \cite{RBF}.   



The rest of the paper is organized as follows. In Sec. \ref{Sec:Model_and_OBF}, the system model is outlined, BS-assisted OBF is reviewed and RS-assisted OBF is introduced. In Sec. \ref{Sec:Asym_Analysis_RS_OBF}, the sum-capacity scaling of RS-assisted OBF is derived for Rayleigh, Rician and correlated Rayleigh fading channels. In Sec. IV, we extend our results to wide-band channels as well as the MISO BC. Simulation results are provided in Sec. \ref{Sec:Simulations} and Sec. VI concludes the paper.

\section{System Model and Opportunistic Beamforming}\label{Sec:Model_and_OBF}

We first outline the downlink transmission model. Then, we review the BS-assisted OBF scheme in \cite{dumb} and introduce the RS-assisted OBF scheme.

\subsection{Transmission Model}

We begin with a downlink network with a single-antenna BS and $K$ single-antenna users. We consider a block-fading model where the channel from BS to user $k\in\{1,\ldots,K\}$, $h_k(t)\in\mathbb{C}$, remains constant during time-slot $t$ of length $T$ samples corresponding to the coherence interval. The received signal $\mathbf{y}_{k}(t) \in \mathbb{C}^{T\times 1}$ at user $k$ in time slot $t$ is 
\begin{align}
\label{t_model}
&\mathbf{y}_{k}(t)=h_{k}(t)\mathbf{s}(t)+\mathbf{n}_{k}(t),
\end{align}
where $\mathbf{s}(t)\in \mathbb{C}^{T\times 1}$ is the vector of Tx symbols, $\mathbf{n}_{k}(t)\in \mathbb{C}^{T\times 1}$ is the Gaussian noise vector at user $k$ with zero mean and identity covariance matrix, i.e., $\mathbf{n}_{k}(t)\sim\mathcal{CN}(\mathbf{0}, \mathbf{I}_{T})$, and $h_k(t)$ is the channel modeled as $h_k(t)=\sqrt{\rho_{B,k}} \tilde{h}_k(t)$, where $\rho_{B,k}$ is the average received SNR\footnote{$\rho_{B,k}$ accounts for the Tx power, channel's large scale fading parameters (path loss, shadow fading), and noise power. The subscript $B$ signifies that $\rho_{B,k}$ is the average SNR of the direct BS-user link.}, while $\tilde{h}_k(t)$ captures the channel's small-scale fading. We assume that the elements of $\mathbf{s}(t)$ are independent and identically distributed (i.i.d.) letters from a Gaussian capacity-achieving codebook with power $\mathbb{E}[\|\mathbf{s}(t)\|^2]=P=1$.


\subsection{Opportunistic Scheduling for Single-Antenna BS}
In such a single-antenna BC with full SNR CSIT, the sum-capacity, defined as the maximum achievable sum of long-term average data rates, is achieved using the OS scheme wherein the BS transmits to the user with the strongest SNR at each fading state \cite{TDMA1, MUdiv1}. The scheduled user in time-slot $t$ is user $\hat{k}(t)= \arg\max_{{k\in\{1,\ldots, K\}}}\gamma_{k}(t)$ where $\gamma_{k}(t) =|h_{k}(t)|^2$ is the SNR of user $k$ at time slot $t$. The maximum SNR is $\gamma_{\hat{k}(t)}(t)$ and the sum-capacity is
\begin{align}
\label{AR}
R^{(K)}= \mathbb{E}[\log_2(1+\max_{k=1,\ldots, K} \gamma_{k}(t))],
\end{align}
where the expectation is over $(h_{1}(t), \ldots, h_{K}(t))$.


In \cite{dumb}, it was shown that when the users' channels undergo independent Rayleigh fading, $R^{(K)}$ increases with $K$, and the performance becomes better than that of the non-fading channel under the same average SNR. This is the consequence of the MU diversity effect, since with larger $K$, it is more likely that a user has SNR much larger than the average SNR. Transmitting to user $\hat{k}(t)$ at time $t$ significantly increases the overall sum-rate. The work in \cite{dumb} studied $R^{(K)}$ in the limit of large $K$ for a homogenous network where $\rho_{B,k}=\rho_B$ $\forall k$ using results from extreme value theory, and showed that under Rayleigh fading channels
\begin{align}
\label{TDMA_ray}
\lim_{K\to \infty } R^{(K)}=\log_2(1+ \rho_{B} \log K).
\end{align}

MU diversity gains are determined by the channel fluctuations' dynamic range. Such gains are often limited due to correlated fading and the presence of LoS channel components, which reduce channel fluctuations and peaks. In fact, when $K$ is large and channels undergo independent Rician fading with parameter $\kappa$ (ratio of energy in the direct component to that in the diffused component), the sum-capacity scales as \cite{dumb}
\begin{align}
\label{TDMA_ric}
&\lim_{K\to\infty } R^{(K)}=\log_2 \hspace{-.1cm}\left(\hspace{-.1cm}1+ \frac{\rho_{B} \left(\sqrt{\log K}+\sqrt{\kappa}\right)^2}{1+\kappa}+O(\log \log K)\hspace{-.1cm}\right)\hspace{-.1cm}.
\end{align}

Compared to Rayleigh fading, the leading term in the SNR of the strongest user is now $\frac{1}{1+\kappa}\log K$ instead of $\log K$, reduced by a factor of $\frac{1}{1+\kappa}$. This is a result of the reduced MU diversity gains due to the presence of the LoS component. In the following subsection, we review an existing scheme that enhances the MU diversity gains in such environments.

\subsection{OBF using Dumb Antennas at the BS}

The authors in \cite{dumb} proposed an OBF scheme that uses $M$ antenna at the BS to induce larger channel fluctuations in environments with dominant LoS channel components or spatial correlation. All antennas transmit the same signal modulated independently by gains that are varied in a controlled but pseudo-random fashion. The resulting overall channel gain in time-slot $t$ becomes
\begin{align}
\label{OBF}
&{h}_{k}(t)=\sqrt{\rho_{B,k}} \sum_{m=1}^M \sqrt{\alpha_{m}(t)} e^{j\theta_{m}(t)} h_{mk}(t),
\end{align}
where $h_{mk}(t)$ is the channel from antenna $m$ to user $k$ at time $t$, $\alpha_m(t)\in[0,1]$ denotes the fraction of power allocated to the $m$-th Tx antenna satisfying $\sum_{m=1}^M \alpha_{m}=1$ to preserve the total power, and $\theta_m(t)\in[0,2\pi]$ denotes the phase shift applied by the antenna. By varying $\alpha_m(t)$ and $\theta_m(t)$, fluctuations in the overall channel can be induced even if $h_{mk}(t)$s have little fluctuations. As in the single-antenna BS case, each user feeds back the overall SNR $|h_{k}(t)|^2$ to the BS, which schedules the user with the largest SNR. There is no need to measure all $h_{mk}(t)$; in fact, the existence of $M$ Tx antennas is completely transparent to the users. This makes the scheme desirable, even though it is not capacity achieving when we have a multi-antenna BS.

The proposed OBF scheme achieves the same performance as \eqref{TDMA_ray} under Rayleigh fading channels. However, when the channels $h_{mk}(t)$s undergo independent Rician fading with parameter $\kappa$, the sum-rate under BS-assisted OBF scales as
\begin{align}
\label{OBF_ric}
&\lim_{K\to\infty } R^{(K)}
=\log_2\hspace{-.1cm} \left(\hspace{-.1cm}1\hspace{-.05cm}+\hspace{-.05cm} \frac{\rho_B(\sqrt{\log K}\hspace{-.05cm}+\hspace{-.05cm}\sqrt{M\kappa})^2}{1+\kappa} \hspace{-.05cm}+\hspace{-.05cm}O(\log \log K)\hspace{-.1cm}\right)\hspace{-.1cm}.
\end{align}
Compared to the single-antenna case, BS-assisted OBF increases the effective magnitude of the fixed component from to $\frac{\kappa}{1+\kappa}$ to $M\frac{\kappa}{1+\kappa}$. While this does not increase the leading term of order $\frac{1}{1+\kappa}\log K$ in the growth rate of the strongest user's SNR, it does increase the second term of order $\sqrt{\log K}$. 

The authors in \cite{dumb} also provide the asymptotic sum-rate under BS-assisted OBF for the special case where $h_{mk}(t)$s are fully correlated, i.e. $h_{mk}(t)=l_{k}(t)e^{j\phi_{mk}}$, where $l_{k}(t)$ is a Rayleigh-fading process and $\phi_{mk}$ are constant phases given as $\phi_{mk}=\phi_{0k}+r(\phi_{0k}, m)$, where $\phi_{0k}$ is the LoS angle to user $k$ and $r$ is a function that abstracts the arrangement of the antenna array. By setting the phases at the BS as $\theta_{m}(t)=-\theta_{0}(t)-r(\theta_{0}(t),m)$ where $\theta_0(t)$ is uniformly rotated, the sum-rate is shown to scale as
\begin{align}
\label{OBF_corr}
\lim_{K\to\infty} R^{(K)}=\log_2 (1+ M \rho_B  \log K).
\end{align}
Thus, BS-assisted OBF yields a factor of $M$ improvement in the SNR of the strongest user as compared to \eqref{TDMA_ray}. 

BS-assisted OBF promises enhanced MU diversity gains using an increased number of antennas at the BS, each of which will have an associated RF chain with several active electronic components (digital-to-analog and analog-to-digital convertors (DAC/ADC), power amplifiers, etc.). This renders the scheme prohibitive from cost and power consumption perspectives, especially when $M$ is large. 


To remedy these challenges, we propose an RS-assisted OBF scheme next, that utilizes a dumb RS to enhance MU diversity gains in different environments, while using a single antenna at the BS and relying on signals reflections from RS elements. 



\subsection{OBF using a Dumb RS}

\begin{figure}
\centering
\includegraphics[scale=.25]{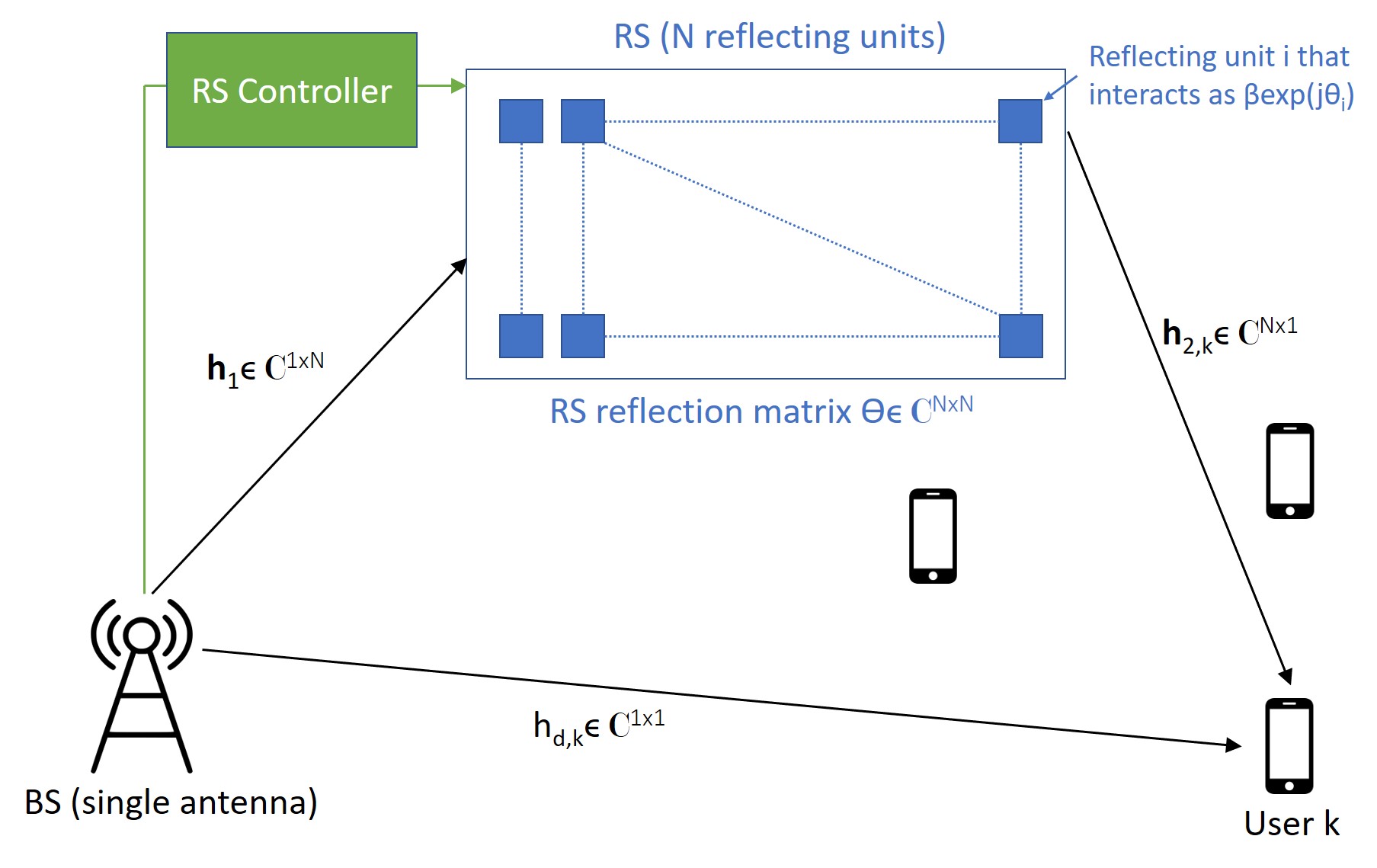}
\caption{RS-assisted OBF system model.}
\label{LIS_model}
\end{figure}

We consider a multi-user communication system where a single-antenna BS serves $K$ single-antenna users, supported by a RS composed of $N$ passive reflecting elements installed in the LoS of the BS (Fig. \ref{LIS_model}). The sum-capacity-achieving transmission strategy is therefore OS. The elements of the RS independently induce random phase shifts onto the incoming waves, rendering the RS to be dumb as it needs no information to design the phases. We assume that the RS's operation is synchronized with the BS through the RS controller and therefore it is able to induce a new set of random phase shifts in each time-slot. The RS-assisted channel gain for user $k$ in time-slot $t$ is given as
\begin{align}
\label{ch_IRS}
h_{R,k}(t)&=\sqrt{\rho_{R,k}} \mathbf{h}_1 \boldsymbol{\Theta}(t) \mathbf{h}_{2,k}(t)\nonumber\\
&=\sqrt{\rho_{R,k}} \mathbf{v}(t)^T \text{diag}(\mathbf{h}_{1}) \mathbf{h}_{2,k}(t),
\end{align}
where $\rho_{R,k}$ is the average SNR of the RS-assisted link, $\mathbf{h}_{1}\in\mathbb{C}^{1\times N}$ is the static LoS BS-RS channel vector \cite{LIS_los, LIS, my_IRS,my_IRS1}, $\mathbf{h}_{2,k}(t) \in \mathbb{C}^{N\times 1}$ is the RS-user $k$ channel vector in time-slot $t$, and $\boldsymbol{\Theta}(t)=\beta\text{diag}\{e^{j\theta_{1}(t)}, \ldots,e^{j\theta_{N}(t)}\} \in \mathbb{C}^{N\times N}$ is a diagonal matrix representing the response of the RS with $\beta$ being the amplitude reflection coefficient and $\theta_n(t)$ the phase shift applied by $n$-th element. The phases $\theta_n(t)$ are varied over time uniformly on $[0,2\pi]$, whereas $\beta$ is fixed\footnote{In practice, $\beta$ is determined by the technology utilized to realize the RS. If reconfigurable metasurface is used, then efficient designs achieving $\beta$ as high as $1$ have been reported \cite{MS_loss, MS_loss2, MS_loss3}.}. The $n$-th component of $\mathbf{h}_1$ is given as $h_{1,n}=e^{j \vartheta_{h_1,n}}$, where $\vartheta_{h_1,n}=2\pi (n-1) d\sin {\vartheta}_n$ \cite{LoS_ent, my_IRS}, ${\vartheta}_n$ is the LoS angle from BS to RS element $n$, and $d$ is the normalized inter-element separation. The second reformulation in \eqref{ch_IRS} has $\mathbf{v}(t)=\beta [e^{j\theta_{1}(t)}, \ldots, e^{j\theta_{N}(t)}]^T$.


Initially, we provide the main theorems for the scenario where the direct BS-user channel is neglected, i.e., the overall channel is $h_k(t)=h_{R,k}(t)$ in \eqref{ch_IRS}. Such a scenario is common in future mmWave and sub-mmWave systems, where the direct BS-user links are likely to be blocked due to high path and penetration losses \cite{spec, LIS1, my_IRS,RIS, RIS1}. This allows us to compare the sum-rate scaling of the RS-assisted link with $N$ passive elements at the RS and a single-antenna BS with that of the direct BS-user link with an $M$-antenna BS. Then, we extend the results to the scenario where the channel is the superposition of the RS-assisted link and the direct link, i.e.,
\begin{align}
\label{ch_ov}
&h_{k}(t)=\sqrt{\rho_{R,k}} \mathbf{h}_1 \boldsymbol{\Theta}(t) \mathbf{h}_{2,k}(t)+ \sqrt{\rho_{B,k}} h_{d,k}(t),
\end{align}
where $h_{d,k}(t)\in \mathbb{C}$ is the small-scale fading channel between the single-antenna BS and user $k$. 

As outlined in Sec. \ref{Sec:Model_and_OBF}, users feedback their SNRs $\gamma_{k}(t)$, and the BS schedules the user with the largest $\gamma_{k}(t)$ in time-slot $t$. Under the proposed RS-assisted OBF model, we will derive the scaling laws of the sum-capacity in \eqref{AR} under 1) independent Rayleigh, 2) independent Rician, and 3) correlated Rayleigh fading $\mathbf{h}_{2,k}(t)$ and $h_{d,k}(t)$, $k=1,\dots,K$. 


\section{Asymptotic Analysis of the Sum-Capacity for RS-Assisted OBF}\label{Sec:Asym_Analysis_RS_OBF}

We derive the sum-capacity scaling laws for RS-assisted OBF and compare them to those of BS-assisted OBF. Similar to \cite{dumb, RBF, slow_f,lit1, lit2, lit3, lit4, lit5, Ric, Ric1,TDMA1}, the analysis is done for a homogenous network where $\rho_{R,k}=\rho_R$ and $\rho_{d,k}=\rho_d$ $\forall k$, to enable the application of extreme value theory results that exist for i.i.d. random variables (RV)s in the literature. However, similar MU diversity effects would be present in a non-homogenous network as we show in Sec. \ref{Sec:Simulations}.\footnote{The downside of OS is that users with higher average SNRs (i.e. lower path loss) will be served more often, making the system unfair. We focus on deriving the fundamental performance limits of RS-assisted systems and do not address the fairness aspect. We refer interested readers to \cite{Ric, Ric1} for algorithms to improve the fairness of systems employing OS (like proportional fair scheduling, user queuing).} 

According to the order statistics for i.i.d. RVs \cite{book_order}, given i.i.d. $\gamma_k$, $k\in\{1,\ldots,K\}$ with a cumulative distribution function (cdf) $F_{\gamma}(x)$ and a probability density function (pdf) $f_{\gamma}(x)$, the cdf and pdf of $\gamma_{\hat{k}}=\max_{k\in\{1,\dots,K\}} \gamma_{k}$ are given as
\begin{align}
\label{1}
F_{\gamma_{\hat{k}}}(x)=[F_{\gamma}(x)]^K,\ f_{\gamma_{\hat{k}}}(x)=K f_{\gamma}(x) [F_{\gamma}(x)]^{K-1},
\end{align}
respectively. The sum-capacity defined in \eqref{AR} is then 
\begin{align}
\label{ex}
R^{(K)}=\int_{0}^{\infty} \log_2(1+x) f_{\gamma_{\hat{k}}}(x) dx.
\end{align}

The resulting expressions using \eqref{ex} for different SNR distributions are involved and yield no direct insights \cite{lit1}. Given the massive connectivity promised by 5G and future generation networks, many systems will have very large $K$. It is thus imperative to study this regime. In what follows, we study the scaling of the sum-capacity versus $K$ for three different fast-fading channel models for $\mathbf{h}_{2,k}$. The derivations will utilize the following result from extreme value theory on the asymptotic behavior of the distribution of the maximum of i.i.d. RVs.

\begin{lemma}{\cite[Lemma 2]{dumb}}\label{Lem:EVT}
Let $x_{1},\dots,x_{K}$ be i.i.d. RVs with pdf $f_{X}(x)$, and cdf $F_{X}(x)$ satisfying $F_{X}(x)<1$ for all finite $x$ and twice differentiable for all $x$. Moreover, let $g(x)=\frac{1-F_{X}(x)}{f_{X}(x)}$ and define $l_K$ as the solution of $F_X(l_{K})=1-\frac{1}{K}$. If 
\begin{align}
\label{growth}
&\underset{x\rightarrow \infty}{\lim}  g(x) =c>0
\end{align}
for some constant $c$, then $\underset{k\in\{1,\ldots,K\}}{\max}x_{k}-l_{K}$ converges in distribution to a limiting RV $z$ with cdf $F_{Z}(z)=e^{-e^{-z/c}}$.
\end{lemma}
This lemma states that the maximum of $K$ i.i.d. RVs described above grows like $l_K$ as $K\to\infty$.

\subsection{Independent Rayleigh Fading}\label{Sec:Asym_Analysis_RS_OBF_Ray}

Dropping the time-index $t$ for simplicity, the RS-assisted channel gain  for user $k$ can be written as $h_{R,k}=\sqrt{\rho_R} \mathbf{v}^T \bar{\mathbf{h}}_{k}$ using \eqref{ch_IRS}, where $\bar{\mathbf{h}}_{k}=[\bar{h}_{k,1},\ldots,\bar{h}_{k,N}]^T=\text{diag}(\mathbf{h}_1) \mathbf{h}_{2,k}$. Under independent Rayleigh faded $\mathbf{h}_{2,k}$s, i.e., $\mathbf{h}_{2,k}\sim \mathcal{CN}(\mathbf{0},\mathbf{I}_{N})$, the $\bar{h}_{k,n}$s are independent circularly symmetric $\mathcal{CN}(0,1)$ RVs since $h_{1,n}h_{1,n}^*=1$ (See Sec. \ref{Sec:Model_and_OBF}-D). Therefore $\mathbf{v}^T \bar{\mathbf{h}}_{k}$ is the sum of $N$ independent Gaussian RVs and is distributed as $\mathcal{CN}(0,\beta^2N)$. The SNR of user $k$  under RS-assisted channel is given as $\gamma_{k}=\rho_R |\mathbf{v}^T \bar{\mathbf{h}}_{k}|^2$ and is distributed as $\gamma_{k}\sim \frac{\rho_R \beta^2 N}{2}  \chi^2(2)$, where  $\chi^2(n)$ is the chi-squared distribution with $n$ degrees of freedom. 


Since $\gamma_{k}$s are i.i.d., we drop the index $k$. The explicit cdf and pdf of $\gamma$ are as follows
\begin{align}
\label{pdf_ray}
F_{\gamma}(x)&=1-e^{-\frac{x}{\rho_R \beta^2 N}},\quad f_{\gamma}(x)=\frac{e^{-\frac{x}{\rho_R \beta^2 N}}}{\rho_R \beta^2 N}, 
\end{align}
respectively, with $x\geq 0$. These distributions can be used to obtain $F_{\gamma_{\hat{k}}}(x)$ and $f_{\gamma_{\hat{k}}}(x)$ using \eqref{1}. The sum-capacity in \eqref{ex} under RS-assisted OBF and Rayleigh fading $\mathbf{h}_{2,k}$s can then be computed as,
\begin{align}
&R^{(K)}\hspace{-.1cm}=\hspace{-.1cm}\int_{0}^{\infty}\hspace{-.1cm}\frac{K \log_2(1+x)}{\rho_R \beta^2 N} e^{-\frac{x}{\rho_R \beta^2 N}}\hspace{-.1cm}\left[1\hspace{-.1cm}-\hspace{-.1cm}e^{-\frac{x}{\rho_R \beta^2 N}}\right]^{K-1}\hspace{-.1cm} dx\nonumber\\
&=\frac{K}{\log 2} \sum_{k=0}^{K-1} \frac{(-1)^k}{k+1} {{K-1}\choose{k}} e^{\frac{k+1}{\rho_R \beta^2 N}} \Gamma\left(0,\frac{k+1}{\rho_R \beta^2 N}\right), \nonumber
\end{align}
where in the last step, we first applied the binomial theorem and then used the identity $\int_{0}^{\infty} e^{-\mu x}\log(1 + \beta x) dx = -\frac{1}{\mu}e^{\frac{\mu}{\beta}}\text{Ei}(-\frac{\mu}{\beta})$, where $\text{Ei}(-x) = -\Gamma(0, x)$ is the exponential integral function, and $\Gamma(a, x)$ is the upper incomplete Gamma function \cite{book_int}.

The resulting closed-form expression for the sum-capacity is not only involved but yields no direct insights. Therefore we study the sum-capacity in the large $K$ regime using Lemma \ref{Lem:EVT}. The result is stated in the following theorem.

\begin{theorem}\label{Thm:RS_OBF_Rayleigh} 
Under independent Rayleigh fading RS-user channels $\mathbf{h}_{2,k}$s, the sum-capacity \eqref{AR} for the channel \eqref{ch_IRS} under RS-assisted OBF scales as,
\begin{align}
\label{Theorem1}
&\underset{K\to \infty}{\lim} R^{(K)}=\log_2 \left(1+ \rho_R \beta^2 N \log K \right).
\end{align} 
\end{theorem}
\begin{IEEEproof}
See Appendix \ref{App:Proof_RS_OBF_Rayleigh}. 
\end{IEEEproof}

\begin{remark} 
As compared to the BS-assisted OBF \cite{dumb}, the RS-assisted OBF improves the SNR of the strongest user by a factor of $N \beta^2$  while using a single antenna at the BS. 
\end{remark}

Theorem \ref{Thm:RS_OBF_Rayleigh} is derived assuming the direct BS-users links are blocked or negligible as compared to the RS-assisted links, as expected in future mmWave and sub-mmWave communications. The extension to the case where the direct link is also present is straightforward and is presented below. 


\begin{corollary}\label{Cor:RS_OBF_Rayleigh}  
Under independent Rayleigh fading RS-user channels $\mathbf{h}_{2,k}$s and BS-user direct channels $h_{d,k}$s, the sum-capacity in \eqref{AR} for the channel in \eqref{ch_ov} under RS-assisted OBF scales as,
\begin{align}
\label{Theorem1_1}
\underset{K\to \infty}{\lim} R^{(K)}\hspace{-.1cm}=\hspace{-.1cm}\log_2 \left(1+ (\rho_R \beta^2 N + \rho_B) \log K \right).
\end{align} 
\end{corollary}
\begin{IEEEproof}
See Appendix \ref{App:Proof_RS_OBF_Rayleigh}.
\end{IEEEproof}

Corollary \ref{Cor:RS_OBF_Rayleigh} simplifies to \eqref{TDMA_ray} when $N=0$ and to \eqref{Theorem1} for large $N$. By increasing $N$ and given $\beta$ is close to unity, the sum-capacity can be made much larger than what is achieved by the conventional single-antenna OS scheme in \eqref{t_model} as well as by the BS-assisted OBF scheme in \eqref{OBF}.

\subsection{Independent Rician Fading}

Rician fading channels generally model situations where there is a fixed LoS component and a time-varying fast fading component. The channel $\mathbf{h}_{2,k}$ under Rician fading is represented as,
\begin{align}
\label{impp}
&\mathbf{h}_{2,k}(t)=\sqrt{a}e^{j \boldsymbol{\phi}_{k}} + \mathbf{b}_{k}(t),
\end{align}
where $a$ is a constant, $\boldsymbol{\phi}_{k}=[\phi_{k,1},\dots, \phi_{k,N}]^T$ is an independent vector of phase shifts of user $k$ LoS channels from the RS, with $\phi_{k,n}$, $n=1,\ldots, N$ representing the phase of the LoS channel component from RS element $n$ to user $k$. Moreover, $\mathbf{b}_{k}=[b_{k,1},\dots, b_{k,N}]^T$ is an independent non LoS fast-fading channel vector from the RS to user $k$, where $b_{k,n}(t)$, $n=1,\ldots, N$, are i.i.d. $\mathcal{CN}(0,u)$ RVs. The $\kappa$-factor is defined as the ratio of the energy in the LoS component to that in the time-varying component, i.e., $a=\frac{\kappa}{1+\kappa}$ and $u=\frac{1}{1+\kappa}$. 

Using \eqref{impp}, the RS-assisted channel $h_{R,k}(t)$ in \eqref{ch_IRS} is written as $h_{R,k}(t)=\sqrt{\rho_{R}}\mathbf{h}_{1}\boldsymbol{\Theta}(t)(\sqrt{a}e^{j \boldsymbol{\phi}_{k}} + \mathbf{b}_{k}(t))$ which can be expanded as,
\begin{align}
\label{sum}
h_{R,k}(t)\hspace{-.5mm}=\hspace{-.5mm}\sqrt{\rho_{R}} \beta\hspace{-.5mm}\sum_{n=1}^N \hspace{-.5mm}h_{1,n}e^{j\theta_{n}(t)}\hspace{-.5mm}\left(\hspace{-.5mm}\sqrt{a} e^{\phi_{k,n}}\hspace{-.5mm} +\hspace{-.5mm} b_{k,n}(t)\right)\hspace{-.5mm}.
\end{align}
Note that $h_{1,n}\beta e^{j\theta_{n}} b_{k,n}$, $n=1,\ldots, N$, are i.i.d. $\mathcal{CN}(0,\beta^2 u)$ RVs. The analysis of the maximum SNR under the RS-assisted channel is not straightforward, since the first component in \eqref{sum} is time-varying. Thus, the users' channels $h_{R,k}$s do not undergo independent Rician fading as $\mathbf{h}_{2,k}$. However, it can be reasoned that for large $K$ and $\epsilon>0$, there exists almost surely at any time $t$ a set of $\epsilon K$ users for which $|\beta\sum_{n=1}^N \sqrt{a} h_{1,n}e^{j (\theta_{n}(t)+\phi_{k,n})}|$ is close to its maximum value $N \sqrt{a} \beta$. Consequently, the overall channel $h_{R,k}(t)$ in \eqref{sum} also undergoes Rician fading for this set of users. The sum-capacity under independent Rician fading can then be studied using Lemma \ref{Lem:EVT} and the result is as follows.

\begin{theorem}\label{Thm:RS_OBF_Rician}
Under independent Rician fading RS-user channels $\mathbf{h}_{2,k}$s, the sum-capacity \eqref{AR} for the channel \eqref{ch_IRS} under RS-assisted OBF scales as,
\begin{align}
\label{Theorem2}
\underset{K\to\infty}{\lim} R^{(K)}&\hspace{-.1cm}=\hspace{-.1cm}\log_2 \left(1+ \frac{\rho_RN\beta^2}{1+\kappa} \left(\sqrt{\log K}+\sqrt{N\kappa}\right)^2 +O(\log \log K)\right).
\end{align}  
\end{theorem}
\begin{IEEEproof} 
See Appendix \ref{App:Proof_RS_OBF_Rician}.
\end{IEEEproof}

Intuitively, this result can be interpreted as: the user with the strongest channel in a time-slot is the one that simultaneously has the strongest time-varying component among all users and has its LoS component in the beamforming configuration, i.e., the norm of $|\beta\sum_{n=1}^N \sqrt{a} h_{1,n}e^{j (\theta_{n}(t)+\phi_{k,n})}|$ is close to $N\beta \sqrt{a}$. Such a user will almost surely exist as $K\to \infty$.


\begin{remark} 
As compared to the sum-rate scaling in \eqref{OBF_ric} for BS-assisted OBF \cite{dumb}, the proposed RS-assisted OBF scheme, with RS elements having $\beta$ close to $1$ \cite{MS_loss, MS_loss2, MS_loss3}, increases the leading term in the growth rate from $u\log K$ to $N\beta^2 u\log K$, the second term from $2\sqrt{Mau\log K}$ to $2N\beta^2\sqrt{Nau\log K}$, and  the fixed component in the growth rate from $Ma$ to $N^2 \beta^2 a$.
\end{remark}

We now extend Theorem \ref{Thm:RS_OBF_Rician} to the scenario where the direct BS-user channel is also present.

\begin{corollary}\label{Cor:RS_OBF_Rician}
Under independent Rician fading RS-user channel vectors $\mathbf{h}_{2,k}$s and BS-user direct channels $h_{d,k}$s, the sum-capacity \eqref{AR} for the channel \eqref{ch_ov} under RS-assisted OBF scales as $\underset{K\to\infty}{\lim} R^{(K)}=\log_2 \left(1+\xi_{\kappa,\kappa_B}^{N,\beta}(K,\rho_R)+O(\log \log K)\right)$ where, 
\begin{align}
\label{Zeta}
\xi_{\kappa,\kappa_B}^{N,\beta}(K,\rho_R)=\left(\sqrt{\left(\frac{\rho_R N\beta^2}{1+\kappa}+\frac{\rho_B }{1+\kappa_{B}}  \right)\log K}+N \beta \sqrt{ \frac{\rho_R\kappa}{1+\kappa}}+ \sqrt{ \frac{\rho_B\kappa_B}{1+\kappa_B}}\right)^2,
\end{align}
where $\kappa_B$ represents the LoS to the non-LoS energy ratio for the BS-to-user direct channels.
\end{corollary}


\subsection{Correlated Rayleigh  Fading}
Now we consider the presence of correlation between the RS elements, i.e., each RS-user channel $\mathbf{h}_{2,k}$ undergoes correlated Rayleigh fading. This correlation is caused by local scatterers around the RS or the fact that the RS elements are not spaced far enough to create independent channels. The overriding question then is to analyze the effect of this correlation on the sum-capacity of RS-assisted OBF. From a traditional diversity and spatial multiplexing point of view, antennas with correlated fading are less useful than antennas with independent fading. From the point of view of OBF, we can actually improve the performance beyond what is achieved by independent fading by exploiting the correlation structure to design OBF parameters.  

%

The channel $\mathbf{h}_{2,k}$ under correlated Rayleigh fading is represented as,
\begin{align}
\label{ch_corr}
&\mathbf{h}_{2,k}(t)=\mathbf{R}_k^{\frac{1}{2}} \mathbf{b}_{k}(t),
\end{align}
where $\mathbf{b}_{k}(t)\sim \mathcal{CN}(\mathbf{0},\mathbf{I}_N)$ and $\mathbf{R}_k$ is the $N\times N$ correlation matrix for user $k$, assumed to be non-singular with $\text{trace}(\mathbf{R}_k) = N$. We consider two scenarios. First we assume a common correlation matrix $\mathbf{R}$ for all $k$. This is essential to make the users' channels (and hence SNRs) i.i.d., enabling the application of results from extreme value theory to get the sum-capacity scaling law. The second scenario deals with completely correlated fading, i.e., the channel coefficients with respect to all RS elements are fully correlated. However the correlation matrix for each user is allowed to be different. 
 

\subsubsection{Common Correlation Matrix}

We first deal with the scenario where $\mathbf{h}_{2,k}(t)=\mathbf{R}^{\frac{1}{2}} \mathbf{b}_{k}(t)$. The overall channel $h_{R,k}(t)$ can be  expressed using \eqref{ch_IRS} as $\sqrt{\rho_R }\mathbf{v}(t)^T \bar{\mathbf{h}}_{k}(t)$, where $\bar{\mathbf{h}}_{k}(t)= \text{diag}(\mathbf{h}_{1}) \mathbf{h}_{2,k}(t)$ and $\mathbf{v}=\beta [e^{j\theta_1}, \ldots, e^{j\theta_N}]^T$.  Therefore, $\bar{\mathbf{h}}_{k}(t)$ is distributed as $\mathcal{CN}(0,\bar{\mathbf{R}})$, where $\bar{\mathbf{R}}=\text{diag}(\mathbf{h}_{1})\mathbf{R}\text{diag}(\mathbf{h}_{1}^H)$, with $\text{trace}(\bar{\mathbf{R}})=N$. The  SNR $\gamma_{k}(t)=|h_{R,k}(t)|^2$ is given as,
\begin{align}
\gamma_{k}(t)&=\rho_R \bar{\mathbf{h}}_{k}(t)^H \bar{\mathbf{v}}(t) \bar{\mathbf{v}}^H(t) \bar{\mathbf{h}}_{k}(t)\\
\label{SNR_corr}
&=\rho \mathbf{z}_{k}(t)^H \mathbf{A}(t)  \mathbf{z}_{k}(t),
\end{align}
where $\mathbf{z}_{k}\sim \mathcal{CN}(\mathbf{0},\mathbf{I}_N)$, $\mathbf{A}(t)=\bar{\mathbf{R}}^{\frac{H}{2}} \bar{\mathbf{v}}(t) \bar{\mathbf{v}}^H(t) \bar{\mathbf{R}}^{\frac{1}{2}}$ and $\bar{\mathbf{v}}(t)=(\mathbf{v}^T(t))^H$.


Finding the distributions of the SNRs is more difficult than for Rayleigh and Rician channel models, as in the latter two scenarios the distribution of the time-varying component did not depend on the distribution of $\mathbf{v}$. However, for correlated Rayleigh channels, the distribution of the SNR in \eqref{SNR_corr} depends on both the distribution of $\mathbf{v}$  as well as the distribution of $\bar{\mathbf{h}}_{k}$. Dropping the time-index for simplicity, we study the sum-capacity in \eqref{AR} as,
\begin{align}
\label{AR2}
R^{(K)}
=\mathbb{E}_{\bar{\mathbf{v}}}[\mathbb{E}_{\bar{\mathbf{h}}_{k}|\bar{\mathbf{v}}}[ \log_2(1+\underset{k}{\max} \gamma_{k})]],
\end{align}
where $\mathbb{E}_{x|y}$ is the conditional expectation of $x$ given $y$. Therefore  we evaluate $R^{(K)}$ by first conditioning on $\mathbf{v}$ and calculating the expectation over $\bar{\mathbf{h}}_{k}$ and then averaging over $\mathbf{v}$. This allows us to study $\underset{k=1,\ldots, K}{\max}\gamma_{k}$ given $\mathbf{v}$ using extreme value theory provided we can evaluate the CDF (and pdf) of $\gamma_{k}$ given $\mathbf{v}$. The sum-capacity scaling for correlated Rayleigh channels under RS-assisted OBF is provided next.

\begin{lemma}\label{Lem:Cor_Rayleigh}
Under correlated Rayleigh fading RS-user channel vectors $\mathbf{h}_{2,k}$s with $\mathbf{R}_k=\mathbf{R}$ $\forall k$, the sum-capacity \eqref{AR} for the channel \eqref{ch_IRS} under RS-assisted OBF scales as,
\begin{align}
\label{AR4}
&R^{(K)}=\underset{\bar{\mathbf{v}}}{\text{max } }\mathbb{E}_{\bar{\mathbf{v}}} [\log_2 (1+ \bar{\mathbf{v}}^H \bar{\mathbf{R}} \bar{\mathbf{v}}\rho_R \log K)].
\end{align}
\end{lemma}
\begin{IEEEproof} 
See Appendix \ref{Proof_Cor_Rayleigh}.
\end{IEEEproof}


The sum-capacity in Lemma \ref{Lem:Cor_Rayleigh} can be derived by designing $\mathbf{v}$ that maximizes the sum-rate scaling under the constraint $|\bar{v}_n|=\beta$ $\forall n$. This is in fact maximized by using a deterministic design for $\mathbf{v}$ that depends on the structure of $\bar{\mathbf{R}}$. The result is stated in the following theorem.

\begin{theorem}\label{Thm:Cor_Rayleigh}
Under the setting of Lemma \ref{Lem:Cor_Rayleigh}, the sum-capacity in \eqref{AR} for RS-assisted OBF scales as,
\begin{align}
\label{AR7}
&R^{(K)}=\log_2(1+ \rho_R N \beta^2 \lambda_{max} \log K),
\end{align}
where $\lambda_{max}$ is the maximum eigenvalue of $\bar{\mathbf{R}}$.
\end{theorem}
\begin{IEEEproof} We exploit the structure of $\bar{\mathbf{R}}$ to employ a deterministic OBF scheme, where $\mathbf{v}$ is fixed over all $t$ and consequently $\bar{\mathbf{v}}$ is fixed. Then \eqref{AR4} can be written as $R^{(K)}=\log_2(1+ \underset{\bar{\mathbf{v}}}{\max} \bar{\mathbf{v}}^H \mathbf{U} \boldsymbol{\Lambda} \mathbf{U}^H \bar{\mathbf{v}} \rho_R  \log K)$, where $\mathbf{U} \boldsymbol{\Lambda} \mathbf{U}^H $ is the eigenvalue decomposition of $\bar{\mathbf{R}}$. The maximum value of $\bar{\mathbf{v}}^H \mathbf{U} \boldsymbol{\Lambda} \mathbf{U}^H \bar{\mathbf{v}}$ under $\|\bar{\mathbf{v}} \|^2=\beta^2 N$ is achieved when $\bar{\mathbf{v}}=\sqrt{N}\beta \mathbf{u}_{max}$, where $\mathbf{u}_{max}$ is the eigen-vector corresponding to the maximum eigenvalue $\lambda_{max}$ of $\bar{\mathbf{R}}$. Thus, the sum-capacity scales as \eqref{AR7}.
\end{IEEEproof}

We also propose a design for $\bar{\mathbf{v}}$ that can achieve the sum-capacity in \eqref{AR7} while satisfying the individual constraints $|\bar{v}_n|=\beta$, $\forall n$. To do this, we set,
\begin{align}
\label{design}
&\bar{\mathbf{v}}=e^{j \angle(\mathbf{u}_{max})}.
\end{align}
 Under this design, $ \bar{\mathbf{v}}^H \mathbf{U} \boldsymbol{\Lambda} \mathbf{U}^H \bar{\mathbf{v}}\simeq N \beta^2 \lambda_{max}$ and therefore the capacity in \eqref{AR7} can be achieved. Since $\text{trace}(\bar{\mathbf{R}})=N$, the maximum eigenvalue $\lambda_{max}\geq 1$ and the term $N \beta^2 \lambda_{max}$ would always be greater than $N \beta^2$. Therefore, in a correlated Rayleigh environment, RS-assisted OBF can  do approximately $\lambda_{max}$ times better in terms of the SNR of the strongest user, than what it attains in independent Rayleigh environment (cf. Theorem \ref{Thm:RS_OBF_Rayleigh}), by exploiting the eigenvalue decomposition of $\bar{\mathbf{R}}$ to design $\mathbf{v}$. We can also reproduce Theorem \ref{Thm:RS_OBF_Rayleigh} as a special case by setting $\mathbf{R}=\mathbf{I}_{N}$ in Lemma \ref{Lem:Cor_Rayleigh}.

 The phase shifts in this design of $\bar{\mathbf{v}}$ are not drawn randomly from the uniform distribution but rather depend on the spatial correlation matrix at the RS. However, the proposed design does not require the RS to be very intelligent since the correlation matrices are well-known to vary very slowly as compared to the fast fading process and therefore only need to be computed after several coherence intervals using information on user's LoS angle and local angular spreads \cite{ourworkTCOM}. This information can be easily acquired by the BS and shared with the RS controller that implements the design of $\bar{\mathbf{v}}$ in \eqref{design}. The RS does not need any information on the fast fading vectors to find the optimal $\bar{\mathbf{v}}$. Moreover, even if we let the phase shifts in $\bar{\mathbf{v}}$ to be drawn randomly from the uniform distribution as done in the rest of the work, the performance of RS-assisted OBF will still be much better than that of a system without the RS as will be shown in Sec. V.

The following corollary extends the results in Theorem \ref{Thm:Cor_Rayleigh} to the scenario where the channel is given by \eqref{ch_ov} and each direct BS-user channel $h_{d,k}$ undergoes independent Rayleigh fading.

\begin{corollary}\label{Cor:RS_OBF_Cor_Rayleigh}
Under correlated Rayleigh fading RS-user channels $\mathbf{h}_{2,k}$s with $\mathbf{R}_k=\mathbf{R}$ and Rayleigh fading BS-user direct channels $h_{d,k}$s, the sum-capacity \eqref{AR} for the channel \eqref{ch_ov} under RS-assisted OBF scales as $\underset{K\to\infty}{\text{lim}} R^{(K)}=\log_2 \left(1+ (\rho_R N\beta^2 \lambda_{max} + \rho_B) \log K \right)$.
\end{corollary}

\subsubsection{Completely Correlated Rayleigh Fading}
Similar to \cite{dumb}, we deal with the scenario where $\mathbf{h}_{2,k}$ is completely correlated, represented as
\begin{align}
\label{ch_c}
\mathbf{h}_{2,k}(t)=l_{k}(t) e^{j \boldsymbol{\phi}_k},
\end{align}
where $l_{k}(t)$ is a Rayleigh-fading process and $\boldsymbol{\phi}_k=[\phi_{1,k},\ldots,\phi_{N,k}]^T$. The phases $\phi_{n,k}$ are constant, and depend on the LoS angle to user $k$ with respect to the RS denoted as $\phi_{0k}$ and the actual placement of the antenna array. For example, for linearly arranged RS elements spaced $d$ units apart (normalized by the wavelength), $\phi_{n,k}=2 \pi d(n-1) \sin \phi_{0k}$. 

The correlation matrix $\mathbf{R}_k$ for $\mathbf{h}_{2,k}$ is then given as $\mathbf{r}_k\mathbf{r}_k^H$ where $\mathbf{r}_k=e^{j \boldsymbol{\phi}_k}$. Also, under this model we have $\bar{\mathbf{R}}_k=\text{diag}(\mathbf{h}_1)\mathbf{R}\text{diag}(\mathbf{h}_1^H)=\bar{\mathbf{r}}_k\bar{\mathbf{r}}_k^H$, where $\bar{\mathbf{r}}_k=e^{j (\boldsymbol{\phi}_k+\boldsymbol{\vartheta}_{h_1})}$, where $\boldsymbol{\vartheta}_{h_1}$ is the vector of phases of $\mathbf{h}_1$, defined in Sec. \ref{Sec:Model_and_OBF}. Note that in this scenario, we allow the correlation matrices to vary across $k$. The overall channel gain of user $k$ is given as $|h_{R,k}(t)|^2=\rho_R \beta^2 \Big|\sum_{n=1}^N e^{j (\theta_{n}(t)+\phi_{n,k}+\vartheta_{h_1,n})}\Big|^2 |l_k(t)|^2$. The SNR is therefore a product of two independent RVs. The maximum value of the first RV is $N^2$, which happens when user $k$ is in the beamforming configuration, i.e. $\theta_{n}(t)+\phi_{n,k}+\vartheta_{h_1,n}=\nu(t) \text{ mod} 2\pi$ $\forall n$, where $\nu(t)$ is an arbitrary phase. Consider a system with large $K$ and with RS phases drawn from the uniform distribution over $[0,2\pi]$ in each time-slot. For any $\delta$, there will almost surely be a fraction $\epsilon \in (0,1)$ of users for which,
\begin{align}
\label{aa}
&\Big|\sum_{n=1}^N e^{j (\theta_{n}(t)+\phi_{n,k}+\vartheta_{h_1,n})}\Big|^2  > N^2 -\delta.
\end{align}

Now $l_{k}$ are i.i.d. Rayleigh distributed random variables - the maximum of which has already shown to scale as $\log K$ in Sec. \ref{Sec:Asym_Analysis_RS_OBF}-B. Hence, among the $\epsilon K$ users in \eqref{aa}, the maximum of $|h_{R,k}(t)|^2$, $k=1,\ldots, K$, grows at least as fast as $\rho_R \beta^2 (N^2-\delta) \log (\epsilon K)=\rho_R \beta^2 (N^2-\delta) \log K + O(1)$ as $K\to \infty$. This is true for every $\delta >0$, and it gives a lower bound to the growth rate of $|h_{R,k}|^2$. Moreover, it is also clear that $\rho_R \beta^2 N^2 \log K$ is an upper bound to the growth rate. Therefore, the sum-capacity scales as stated in the following corollary.

\begin{corollary} \label{Cor:RS_OBF_Comp_Cor_Rayleigh}
When the RS-user channel vectors $\mathbf{h}_{2,k}$s independently undergo completely correlated Rayleigh fading as described in \eqref{ch_c}, the sum-capacity in \eqref{AR} for RS-assisted OBF scales as,
\begin{align}
\label{Cor4}
R^{(K)}=\log_2(1+ \rho_R N^2 \beta^2 \log K).
\end{align}
\end{corollary}

The result can also be derived directly from Theorem \ref{Thm:Cor_Rayleigh} by noting that for a rank-one $\mathbf{R}$, the maximum eigenvalue $\lambda_{max}=N$. However, the analysis in this section shows that the sum-rate scaling in Theorem \ref{Thm:Cor_Rayleigh} for a common correlation matrix for all $k$ extends to per-user correlation matrices as well.

\begin{remark} 
As compared to the sum-rate scaling given in \eqref{OBF_corr} under BS-assisted OBF \cite{dumb}, the RS-assisted OBF yields a factor of $N$ improvement in the SNR of the strongest user. 
\end{remark}


\begin{remark} 
As compared to the sum-rate scaling in Theorem \ref{Thm:RS_OBF_Rayleigh} for independent Rayleigh fading under RS-assisted OBF, we get a factor of $N$ improvement in the SNR of the strongest user under completely correlated Rayleigh model. This is because the overall channel is no longer Rayleigh; instead, it is a mixture of Gaussian distributions with different variances. When the received signals from the RS elements add in phase, the overall received SNR is large; when they add out of phase, the overall SNR is small. This yields additional MU-diversity gain.
\end{remark} 


\section{Extension to Wide-Band Channels and Multi-Antenna Systems}

In this section, we extend the RS-assisted OBF scheme to OFDMA and multi-antenna systems. 

\subsection{Wide-band Channels}
In wide-band downlink channels, OFDMA is an appropriate transmission protocol which allows serving multiple users simultaneously by scheduling them to orthogonal frequency bands. The works proposing BS-assisted OBF in OFDMA systems exploit MU diversity in the frequency domain in addition to the time domain, i.e., users are scheduled on their frequency fading peaks as well \cite{OBFOFDMA, OBFOFDMA1}. While MU diversity gain in a single-carrier system is obtained by scheduling the best user (in terms of SNR) for transmission in a time slot, MU-diversity in frequency-selective fading channels is obtained by scheduling the best user per frequency band for transmission on that frequency band in a time-slot. All users share all bands, translating into $K$ users per band for OBF to capitalize on. RS-assisted OBF can yield significant gains in such systems. 

A simple model of a frequency-selective wide-band channel is a set of $L$ parallel narrow-band sub-channels. An $L$ sub-carrier OFDMA system with proper cyclic prefix extension converts the frequency selective channel into $L$ flat-fading sub-channels, where the $l$-th sub-channel between the BS and user $k$ under RS-assisted OBF is given as,
\begin{align}
\label{ch_ov1}
&h_{k}^{(l)}(t)\hspace{-.09cm}=\hspace{-.09cm}\sqrt{\hspace{-.07cm}\rho^{(l)}_{R,k}} \mathbf{h}^{(l)}_1 \boldsymbol{\Theta}(t) \mathbf{h}^{(l)}_{2,k}(t)\hspace{-.09cm}+\hspace{-.09cm} \sqrt{\hspace{-.07cm}\rho^{l}_{B,k}} h^{(l)}_{d,k}(t),
\end{align}
for $l\in\{1,\ldots,L\}$. The users measure the SNR on each of the sub-channels and feed back the SNRs to the BS. Thus, this requires $L$ times more feedback than in the flat-fading case where only a single SNR is fed back by each user. The BS transmits to the best user per sub-channel at each time slot. Thus, the sum-capacity per sub-channel scales exactly as it does for the single-carrier system and the total sum-capacity is just the sum over all $L$ sub-channels. Assuming a homogenous network, we summarize the asymptotic sum-capacity scaling as follows.

\begin{corollary} \label{Cor:RS_OBF_Rayleigh_OFDMA}
In the frequency-selective wide-band channel, where $\mathbf{h}^{(l)}_{2,k}$ and $h^{(l)}_{d,k}$ undergo independent Rayleigh fading, the sum-capacity using an $L$ sub-carrier OFDMA system scales as $\underset{K\to\infty}{\lim} R^{(K)}=\sum_{l=1}^L \log_2 \left(1+ (\rho^{(l)}_R \beta^2 N + \rho^{(l)}_B) \log K \right)$.
\end{corollary}

\begin{corollary} \label{Cor:RS_OBF_Rician_OFDMA}
In the frequency-selective wide-band channel, where $\mathbf{h}^{(l)}_{2,k}$ and $h^{(l)}_{d,k}$ undergo independent Rician fading, the sum-capacity using an $L$ sub-carrier OFDMA system scales as $\underset{K\to\infty}{\lim} R^{(K)}=\sum_{l=1}^L \log_2 \left(1 +\xi_{\kappa^{(l)},\kappa_B^{(l)}}^{N,\beta}(K,\rho_R^{(l)}) +O(\log\log K)\right)$, where $\xi_{\kappa^{(l)},\kappa_B^{(l)}}^{N,\beta}(K,\rho_R^{(l)})$ is defined in \eqref{Zeta}.
\end{corollary}

The advantage of having many sub-channels shared by $K$ users is reflected by an $L$ fold increased scaling, since the MU-diversity effect is exploited in both time as well as frequency domain. 

\subsection{Multiple Antennas at the BS}\label{Sec:MISO_BS}


We extend our results to the MISO BC under the OS strategy, which is no longer optimal, since instead of sending a single beam to the best user, multiple beams can be sent in the same coherence interval to serve multiple users using beamforming. Such a beamforming scheme combined with DPC is capacity-optimal for the MISO BC, but it requires a large amount of feedback as the BS needs perfect CSIT for all the users with respect to all the Tx antennas. OS, on the other hand, requires just one SNR feedback per user irrespective of the number of antennas and is simple to implement. Thus, like \cite{dumb,Ric1,Ric, slow_f,OBFOFDMA}, we continue to focus on OS even for the MISO BC and study the impact of RS-assisted OBF on the sum-rate. 

Consider a BS is equipped with $M$ antennas. The received signal $y_k(t)$ at user $k$ in time-slot $t$ is given as $y_k(t)=\mathbf{h}_k^H(t) \mathbf{s}(t) + n_k(t)$, where $\mathbf{h}_k^H(t)\in \mathbb{C}^{1\times M}$ is the downlink channel from BS to user $k$ and $n_k(t)$ is noise. Moreover $\mathbf{s}\in \mathbb{C}^{M\times 1}$ is the Tx signal vector assumed  to have i.i.d. Gaussian entries with zero mean and unit variance. This maintains the average Tx power per antenna as unity, similar to the MISO BC scenario considered in \cite{dumb}, i.e., $\mathbb{E}[\mathbf{s}(t)\mathbf{s}^H(t)]=\mathbf{I}_M$ and $\mathbb{E}[\|\mathbf{s}(t)\|^2]=M$.

As in the single Tx antenna case, each user will just feed back the overall SNR $\gamma_k(t)=\|\mathbf{h}_{k}(t)\|^2$ to the BS, which will schedule according to OS. The average achievable sum-rate is given by \eqref{AR}. This setting is exactly the BS-assisted OBF scenario discussed in Sec. II-C \cite{dumb, Ric1}, for which the sum-rate scaling under Rayleigh and Rician fading $\mathbf{h}_k$s is given by \eqref{TDMA_ray} and \eqref{OBF_ric}, respectively. The use of multiple antennas under BS-assisted OBF yields no gain in Rayleigh fading channels but it provides gain in Rician fading environments due to the increase in MU-diversity. 

We now study the sum-rate scaling for RS-assisted OBF in a MISO BC. The channel \eqref{ch_IRS} is expressed as $\mathbf{h}_{R,k}(t)=\sqrt{\rho_{R}} \mathbf{H}_1 \boldsymbol{\Theta}(t) \mathbf{h}_{2,k}(t)$, where $\mathbf{H}_1$ is an $M\times N$ LoS channel matrix with complex exponential components, i.e. $[\mathbf{H}_1]_{m,n}=e^{j 2\pi (m-1) \sin(\vartheta_n')}$, $m=1,\ldots, M$, $n=1,\ldots, N$. When $\mathbf{h}_{2,k}(t)$ undergoes Rayleigh fading, i.e., $\mathbf{h}_{2,k}\sim \mathcal{CN}(\mathbf{0}, \mathbf{I}_N)$, the overall channel $\mathbf{h}_{R,k}(t)$ undergoes correlated Rayleigh fading with correlation matrix $\mathbb{E}[\mathbf{h}_{R,k}(t)\mathbf{h}_{R,k}(t)^H]=\rho_{R} \beta^2 \mathbf{H}_1\mathbf{H}_1^H$. 
Let us define $\mathbf{R}=\frac{1}{N}\mathbf{H}_1\mathbf{H}_1^H$, so that $\text{tr}(\mathbf{R})=M$, and write $\mathbf{h}_{R,k}$ (dropping $t$) as,
\begin{align}
\label{ch_corr1}
&\mathbf{h}_{R,k}=\sqrt{\rho_{R} N} \beta \mathbf{R}^{\frac{1}{2}} \mathbf{z}_k,
\end{align} 
where $\mathbf{z}_k\sim \mathcal{CN}(\mathbf{0}, \mathbf{I}_M)$. The analysis developed for the MISO BC under BS-assisted OBF in \cite{Ric1, dumb}  can not be directly applied here as $\mathbf{h}_{R,k}$ undergoes correlated Rayleigh fading. Developing the pdf/cdf expressions of the SNR $\gamma_k(t)=\|\mathbf{h}_{R,k}(t)\|^2$ to exploit the results in Lemma \ref{Lem:EVT} is quite involved and out of the scope of this work. However, we can use the channel whitening idea in \cite{tareq} to convert the correlated Rayleigh channel \eqref{ch_corr1} to a Rayleigh fading channel and obtain the sum-rate scaling using Lemma \ref{Lem:EVT}.\footnote{The authors in \cite{tareq} used channel whitening to study the sum-rate scaling of random beamforming in correlated Rayleigh channels. The sum-rate scaling using the true cdf/pdf of the SNR was also derived and simulation results showed the sum-rate obtained using channel whitening to be quite close to the true values.}  

To whiten the channel, we multiply $\mathbf{s}$ with $\zeta \mathbf{R}^{-\frac{1}{2}}$ where $\zeta$ is a normalization factor. The new transmit signal $\mathbf{x}\in \mathbb{C}^{M\times 1}$ is therefore equal to $\mathbf{x}=\sqrt{\zeta} \mathbf{R}^{-\frac{1}{2}} \mathbf{s}$, where $\zeta$ is chosen to satisfy the Tx power constraint $\mathbb{E}[\mathbf{x}^H\mathbf{x}]=\mathbb{E}[\zeta \mathbf{s}^H \mathbf{R}^{-1} \mathbf{s}]=M$ and can be shown to be $\zeta=\frac{M}{\text{tr }(\mathbf{R}^{-1})}$. The received SNR is thus 
\begin{align}
\gamma_k&=\zeta \rho_R \beta^2 N \mathbf{z}_k^H \mathbf{R}^{\frac{1}{2}}\mathbf{R}^{-\frac{1}{2}}\mathbb{E}[\mathbf{s}\mathbf{s}^H] \mathbf{R}^{-\frac{1}{2}}\mathbf{R}^{\frac{1}{2}} \mathbf{z}_k\\
&=\zeta \rho_R \beta^2 N \mathbf{z}_k^H  \mathbf{z}_k.
\end{align}

The SNR of each user is distributed as $\frac{\zeta \rho_R \beta^2 N}{2} \chi^2(2M)$ and the resulting cdf and pdf are, 
\begin{align}
F_{\gamma}(x)\hspace{-.5mm}=\hspace{-.5mm}1-e^{-\frac{x}{c}}\hspace{-.7mm}\sum_{m=0}^{M-1} \hspace{-.7mm}\frac{x^m}{c^m m!},\ \ f_{\gamma}(x)\hspace{-.5mm}=\hspace{-.5mm}\frac{x^{M-1} e^{-\frac{x}{c}}}{c^M(M-1)! },  \nonumber
\end{align}
respectively, where $c=\zeta \rho_R \beta^2 N$. Using Lemma \ref{Lem:EVT}, we obtain $g(x)=\frac{1-F_{\gamma}(x)}{f_{\gamma}(x)} =\zeta \rho_R \beta^2 N >0$, thereby satisfying \eqref{growth}. Therefore $\gamma_{\hat{k}}-l_{K}$ converges in distribution to a limiting RV $z$. Solving for $l_K$ from Lemma \ref{Lem:EVT}, we obtain $l_{K}=\zeta \rho_R \beta^2 N \log K + O(\log \log K)$. Plugging this for $\underset{k=1,\ldots, K}{\max} \gamma_k$ in \eqref{AR} yields the following sum-rate scaling. 


\begin{theorem}\label{Thm:RS_OBF_Rayleigh_MISOBC}
Under independent Rayleigh fading RS-user channel vectors $\mathbf{h}_{2,k}$s, the sum-rate \eqref{AR} for the channel \eqref{ch_corr1} under RS-assisted OBF and channel whitening scales as,
\begin{align}
\label{Theorem4}
\underset{K\to\infty}{\lim} R^{(K)}&=\log_2 \left( \rho_R \beta^2 MN \log K + O(\log \log K)\right) - \log_2(\text{tr }(\mathbf{R}^{-1})).
\end{align} 
\end{theorem}


Theorem \ref{Thm:RS_OBF_Rayleigh_MISOBC} yields some important insights. First, RS-assisted OBF in the MISO BC improves the SNR of the strongest user by a factor of $MN$ compared to the single-antenna BC under OS in \eqref{TDMA_ray}, but also introduces a loss in the sum-rate due to the channel correlation which arises from the fixed BS-RS LoS channel matrix $\mathbf{H}_1$. On the other hand, the BS-assisted OBF system discussed in Sec. \ref{Sec:Model_and_OBF}-C yields no gain through the use of multiple antennas at the BS in a Rayleigh fading environment. Second, as opposed to the single-antenna RS-assisted OBF scheme in Sec. \ref{Sec:Asym_Analysis_RS_OBF_Ray}, there is a factor-$M$ gain in the SNR of the strongest user and a loss of $\log_2\text{tr}\left(\mathbf{R}^{-1}\right)$ in performance. In fact, the gain achieved by RS-assisted OBF with a multi-antenna BS can easily be achieved by having only a single-antenna BS and using a higher number of passive reflecting elements at the RS. In fact, since only one user is served in each time-slot, it does not benefit much to have multiple active antennas at the BS especially in terms of energy efficiency. Similar analysis applies for the scenario where $\mathbf{h}_{2,k}$ undergo Rician fading and correlated Rayleigh fading. 

\begin{remark} \label{Rem:Scaling_Power_MISO_BC}
Many works assume the total transmit power $P$ to be equal to one (and not scale with $M$), such that $\mathbb{E}[\|\mathbf{s}(t)\|^2]=1$ \cite{Ric, Ric1},  which implies that $\mathbb{E}[\mathbf{s}(t)\mathbf{s}^H(t)]=\frac{1}{M}\mathbf{I}_M$. Under this constraint, the sum-rate scaling in \eqref{Theorem4} is given as ${\lim}_{K\to \infty} R^{(K)}=\log_2 ( \rho_R \beta^2 N \log K +\\ O(\log \log K)) - \log_2 \text{tr } (\mathbf{R}^{-1})$. In this case, the MISO system performs worse than the single antenna system under RS-assisted OBF, since there is no array gain due to the use of multiple antennas at the BS but rather there is a loss of $\log_2 \text{tr }(\mathbf{R}^{-1})$. 
\end{remark}

As an extension of this work, it is interesting to study the impact of introducing a dumb RS in the MISO BC under the SDMA strategy, where multiple users are served in the same time slot using beamforming. A popular scheme that can be studied is random beamforming proposed in \cite{RBF}, where $M$ orthogonal beams are constructed at the BS and on each beam the best user is served. This scheme is an extension of BS-assisted OBF and also capitalizes on MU-diversity gains. Studying the impact of deploying an RS on the sum-rate scaling of the RBF scheme is an interesting extension and is left for future work.

\section{Simulations}\label{Sec:Simulations}

This section presents simulation results that verify the derived sum-rate scaling laws. We first compare the proposed RS-assisted OBF scheme for a single-antenna BS supported by a dumb RS with $N$ elements against the BS-assisted OBF scheme for an $M$-antenna BS. The channel model for the former is given by \eqref{ch_ov} (both RS-assisted and direct links are considered) and for the latter is given by \eqref{OBF}. We set the reflection coefficient $\beta$ to $1$ and the average SNR for all links to $0$ $\rm{dB}$, unless otherwise stated.



\begin{figure*}[!t]
\begin{subfigure}[t]{.325\textwidth}
\tikzset{every picture/.style={scale=.95}, every node/.style={scale=.8}}
\input{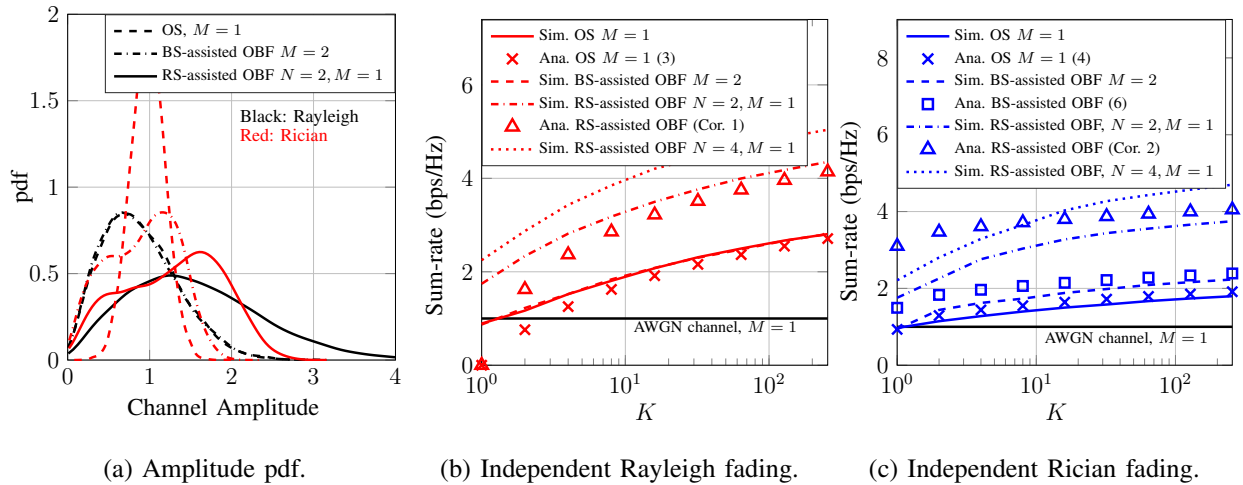}
\caption{Amplitude pdf.}
\label{Fig1_1}
\end{subfigure}
\begin{subfigure}[t]{.325\textwidth}
\tikzset{every picture/.style={scale=.95}, every node/.style={scale=.8}}
%
%
\definecolor{mycolor1}{rgb}{0.63529,0.07843,0.18431}%
\definecolor{mycolor2}{rgb}{0.00000,0.49804,0.00000}%
\definecolor{mycolor3}{rgb}{0.92941,0.69412,0.12549}%
\definecolor{mycolor4}{rgb}{1.00000,0.00000,1.00000}%
\begin{tikzpicture}

\begin{axis}[%
width=.95\columnwidth,
height=.95\columnwidth,
scale only axis,
xmin=1,
xmax=256,
xmode=log,
xlabel={$K$},
xmajorgrids,
ymin=0,
ymax=7.4,
ylabel style = {at={(axis cs: 1.5,3)}},
ylabel={Sum-rate (bps/Hz)},
ymajorgrids,
legend style={at={(axis cs: 1,7.4)},anchor=north west,legend cell align=left,align=left,draw=white!15!black, /tikz/column 2/.style={
                column sep=5pt,
            }},
]
\addplot [color=black,solid,line width=1.0pt, forget plot]
  table[row sep=crcr]{%
1	1\\
2	1\\
4	1\\
8	1\\
16	1\\
32	1\\
64	1\\
128	1\\
256	1\\
};
\node at (axis cs: 10,.8) [anchor = west] {\scriptsize AWGN channel, $M=1$};

\addplot [color=red,solid,line width=1.0pt]
  table[row sep=crcr]{%
1	0.892014794604175\\
2	1.16881835665892\\
4	1.52115541977843\\
8	1.80604321258485\\
16	2.07618528252656\\
32	2.31065385231972\\
64	2.49757189440943\\
128	2.66874894548746\\
256	2.806528328614\\
};
\addlegendentry{\scriptsize Sim. OS $M=1$};

\addplot [color=red,only marks ,line width=1.0pt,mark size=3.0pt,mark=x,mark options={solid}]
  table[row sep=crcr]{%
1	0\\
2	0.759707388138909\\
4	1.25477201752352\\
8	1.62266874114734\\
16	1.91555482987859\\
32	2.15889793477275\\
64	2.36705875081923\\
128	2.54893723042887\\
256	2.71043230520344\\
};
\addlegendentry{\scriptsize Ana. OS $M=1$ \eqref{TDMA_ray}};

\addplot [color=red,dashed,line width=1.0pt]
  table[row sep=crcr]{%
1	0.870978247644017\\
2	1.22512527906862\\
4	1.52302196547093\\
8	1.83686023658686\\
16	2.08316423753276\\
32	2.30127768317016\\
64	2.47944974051709\\
128	2.66405297910488\\
256	2.80725909447864\\
};
\addlegendentry{\scriptsize Sim. BS-assisted OBF $M=2$};

\addplot [color=red,dash dot,line width=1.0pt]
  table[row sep=crcr]{%
1	1.74315854001439\\
2	2.3375062019243\\
4	2.81683229947799\\
8	3.18130749901208\\
16	3.49727752960996\\
32	3.76218658843746\\
64	3.99721816716999\\
128	4.17764771367117\\
256	4.35082368516847\\
};
\addlegendentry{\scriptsize Sim. RS-assisted OBF $N=2,M=1$};

\addplot [color=red,only marks,line width=1.0pt,mark size=3.0pt,mark=triangle,mark options={solid}]
  table[row sep=crcr]{%
1	0\\
2	1.62266874114734\\
4	2.36705875081923\\
8	2.85565581127947\\
16	3.21998412585876\\
32	3.51060850535784\\
64	3.7523899333058\\
128	3.9594076549597\\
256	4.14041321945906\\
};
\addlegendentry{\scriptsize Ana. RS-assisted OBF (Cor. \ref{Cor:RS_OBF_Rayleigh})};

\addplot [color=red,dotted,line width=1.0pt]
  table[row sep=crcr]{%
1	2.24846476355725\\
2	2.8720852541135\\
4	3.43148269228614\\
8	3.8482489144755\\
16	4.20509195654298\\
32	4.43746729311503\\
64	4.67483352648657\\
128	4.87205934347051\\
256	5.03973288143191\\
};
\addlegendentry{\scriptsize Sim. RS-assisted OBF $N=4,M=1$};
\end{axis}
\end{tikzpicture}%
\caption{Independent Rayleigh fading.}
\label{Fig1}
\end{subfigure}
\begin{subfigure}[t]{.325\textwidth}
\tikzset{every picture/.style={scale=.95}, every node/.style={scale=.8}}
%
%
\definecolor{mycolor1}{rgb}{0.63529,0.07843,0.18431}%
\definecolor{mycolor2}{rgb}{0.00000,0.49804,0.00000}%
\definecolor{mycolor3}{rgb}{0.92941,0.69412,0.12549}%
\definecolor{mycolor4}{rgb}{1.00000,0.00000,1.00000}%
\begin{tikzpicture}

\begin{axis}[%
width=.92\columnwidth,
height=.95\columnwidth,
scale only axis,
xmin=1,
xmax=256,
xmode=log,
xlabel={$K$},
xmajorgrids,
ymin=0,
ymax=9,
ylabel style = {at={(axis cs: 1.5,4)}},
ylabel={Sum-rate (bps/Hz)},
ymajorgrids,
legend style={at={(axis cs: 1,9)},anchor=north west,legend cell align=left,align=left,draw=white!15!black, /tikz/column 2/.style={
                column sep=5pt,
            }},
]
\addplot [color=black,solid,line width=1.0pt,forget plot]
  table[row sep=crcr]{%
1	1\\
2	1\\
4	1\\
8	1\\
16	1\\
32	1\\
64	1\\
128	1\\
256	1\\
};
\node at (axis cs: 10,0.7) [anchor = west] {\scriptsize AWGN channel, $M=1$};

\addplot [color=blue,solid,line width=1.0pt]
  table[row sep=crcr]{%
1	0.969880538345253\\
2	1.14561473105942\\
4	1.27806101939451\\
8	1.39736962559514\\
16	1.49990333415012\\
32	1.58030373133603\\
64	1.6692866125601\\
128	1.73783624440205\\
256	1.79641957195297\\
};
\addlegendentry{\scriptsize Sim. OS $M=1$};

\addplot [color=blue,only marks,line width=1.0pt,mark size=3.0pt,mark=x,mark options={solid} ]
  table[row sep=crcr]{%
1	0.932885804141463\\
2	1.29324659557733\\
4	1.43940017931665\\
8	1.54954015499867\\
16	1.64085916097406\\
32	1.72007290796087\\
64	1.79064935651302\\
128	1.8546606437925\\
256	1.9134636298675\\
};
\addlegendentry{\scriptsize Ana. OS $M=1$ \eqref{TDMA_ric}};

\addplot [color=blue,dashed, line width=1.0pt]
  table[row sep=crcr]{%
1	0.938513386194658\\
2	1.43848529026249\\
4	1.62196362913295\\
8	1.73315556551054\\
16	1.88183858989534\\
32	1.98340352395393\\
64	2.08901845417351\\
128	2.16215864280555\\
256	2.23154917340349\\
};
\addlegendentry{\scriptsize Sim. BS-assisted OBF $M=2$};

\addplot [color=blue,only marks, line width=1.0pt,mark size=2.0pt,mark=square,mark options={solid}]
  table[row sep=crcr]{%
1	1.49476469174958\\
2	1.83113072243748\\
4	1.96405813757418\\
8	2.06341891705438\\
16	2.14540410613849\\
32	2.21629479397219\\
64	2.2793157046983\\
128	2.33638384781893\\
256	2.38874882478106\\
};
\addlegendentry{\scriptsize Ana. BS-assisted OBF \eqref{OBF_ric}};

\addplot [color=blue,dash dot,line width=1.0pt]
  table[row sep=crcr]{%
1	1.75590574364031\\
2	2.25770565822944\\
4	2.75880801090544\\
8	3.0360039175103\\
16	3.2741966587137\\
32	3.43830761406988\\
64	3.55529009325536\\
128	3.65856725393378\\
256	3.74728695323228\\
};
\addlegendentry{\scriptsize Sim. RS-assisted OBF, $N=2,M=1$};

\addplot [color=blue,only marks,line width=1.0pt,mark size=3.0pt,mark=triangle,mark options={solid}]
  table[row sep=crcr]{%
1	3.0987798641145\\
2	3.46781067395453\\
4	3.60901511371466\\
8	3.71322768597319\\
16	3.79846200978661\\
32	3.8716666735989\\
64	3.93639145798883\\
128	3.99473714798567\\
256	4.04806748636666\\
};
\addlegendentry{\scriptsize Ana. RS-assisted OBF (Cor. \ref{Cor:RS_OBF_Rician})};

\addplot [color=blue,dotted,line width=1.0pt]
  table[row sep=crcr]{%
1	2.21107499321167\\
2	2.80377321104179\\
4	3.28873471971672\\
8	3.6499241536822\\
16	4.02914734722003\\
32	4.27004920711074\\
64	4.42929707485624\\
128	4.5680887524023\\
256	4.69855747267872\\
};
\addlegendentry{\scriptsize Sim. RS-assisted OBF, $N=4,M=1$};
\end{axis}
\end{tikzpicture}%
\caption{Independent Rician fading.}
\label{Fig1_2}
\end{subfigure}
\caption{Channel amplitude pdf and simulated (Sim.) and analytical (Ana.) sum-rate for BS-assisted and RS-assisted OBF.}
\label{Fig1_all}
\end{figure*}

Fig. \ref{Fig1_1} shows the pdf of the amplitude of the overall channel $|h_k(t)|$ under Rayleigh and Rician fading and BS-assisted and RS-assisted OBF. The parameter $\kappa=\kappa_B$ is set as $10$ for Rician fading. It can be seen that $M$ has no effect on the overall channel amplitude in a Rayleigh faded channel under BS-assisted OBF. However, RS-assisted OBF  increases the dynamic range of channel fluctuation in a Rayleigh environment from $3$ to $5$. Therefore, we should expect larger MU-diversity gains with RS-assisted OBF. In contrast, in Rician fading channels, BS-assisted OBF does increase the the overall channel  amplitude by increasing the norm of the fixed component. However, the increase is more with RS-assisted OBF. This is in accordance with our results in Theorem \ref{Thm:RS_OBF_Rician} and Corollary \ref{Cor:RS_OBF_Rician}, where we showed  that the RS increases the fluctuations of both the fixed and time-varying channel components under Rician fading.  


Fig. \ref{Fig1} and Fig. \ref{Fig1_2} study the relation between the sum-rate in \eqref{AR} and $K$ for Rayleigh and Rician channels under BS-assisted and RS-assisted OBF. The analytical expressions of the scaling laws in \eqref{TDMA_ray} and \eqref{TDMA_ric} for the conventional single-antenna system under OS, \eqref{TDMA_ray} and \eqref{OBF_ric} for BS-assisted OBF, and in Corollary (Cor.) \ref{Cor:RS_OBF_Rayleigh} and \ref{Cor:RS_OBF_Rician} for RS-assisted OBF are also plotted. These figures confirm that the sum-rates are increasing with $K$ as a result of MU-diversity gains. The simulated values approach the derived scaling laws for large $K$. The convergence is slower for Rician fading because the proof requires the best user to be the one that simultaneously maximizes the norm of the fixed component as well as the time-varying component, which requires a much larger $K$. The pdf curves in Fig. \ref{Fig1_1} showed larger channel amplitudes for Rayleigh than Rician fading channels. Therefore OBF yields higher sum-rate in Rayleigh fading channels. 

	

As discussed in Sec. \ref{Sec:Model_and_OBF}-C, BS-assisted OBF does not help under Rayleigh fading channels. RS-assisted OBF with $N=2$ elements, however, yields approximately $\log_2 (N+1)=\log_2(3)=1.5$ bits/transmission improvement under Rayleigh fading channels,  as shown in Fig. \ref{Fig1} which is in line with Corollary \ref{Cor:RS_OBF_Rayleigh}. Under Rician fading, BS-assisted OBF increases the overall channel magnitude resulting in an improvement in the sum-rate of approximately $.4$ bits/transmission as shown in Fig. \ref{Fig1_2}, corresponding to the difference between \eqref{TDMA_ric} and \eqref{OBF_ric}. RS-assisted OBF further increases the sum-rate due to the larger increase in the channel amplitude illustrated in Fig. \ref{Fig1_1}. Compared to BS-assisted OBF, we see approximately $\log_2 (N+1)$ improvement in the sum-rate achieved by RS-assisted OBF under Rician fading channels in Fig. \ref{Fig1_2}, which is in line with the sum-rate scaling in Corollary \ref{Cor:RS_OBF_Rician}. 



The previous results assumed a homogenous network where the average received SNR of each user is the same (i.e. $\rho_{B,k}=\rho_{B},$ $\rho_{R,k}=\rho_R$, $\forall k$) similar to the theoretical analysis in Sec. \ref{Sec:Asym_Analysis_RS_OBF}.  However, even with per-user path loss, similar MU-diversity gains can be realized. Next, we simulate the sum-rate for a non-homogenous network.

It has been shown that if the size of each meta-surface (element) of the RS is large enough compared to the wavelength $\lambda$ of the radio wave (e.g. $10\lambda$ \cite{spec, spec1}), then each metasurface behaves approximately as an anomalous (specular) reflector. In this case, the power received from each metasurface scales as $(d_{B-R}+d_{R-U})^2$ and not as $(d_{B-R}d_{R-U})^2$, where $d_{B-R}$ and $d_{R-U}$ are the BS-RS and RS-user distances, respectively. In fact, \cite{spec} shows that the RS composed of $N= 100$ reconfigurable meta-surfaces acting as anomalous reflectors would be of the order of $100\lambda\times 100\lambda$ in size. If  the  operating  frequency  is  $30GHz$ (mm-Wave  frequency),  then the resulting  RS is of area $1  m^2$. Such a RS can be readily deployed which means we can assume the path loss to scale as $\lambda(4\pi(d_{B-R}+d_{R-U}))^{-2}$. Similar path loss is assumed for BS-user link and the distance is represented as $d_{B-U}$. Using $(x,y,z)$ coordinates (in meters), the BS and RS are deployed at $(0,0,0)$ and $(50,50,40)$, respectively, and the users are assumed to be uniformly distributed in the square $(x,y)\in[0,100]\times[50,150]$. We assume an operating frequency of $28 GHz$ and bandwidth of $100 MHz$. Due to the fact that $d_{B-R}+d_{R-U}>d_{B-U}$, we expect that RS performance will suffer, and compared to Fig. \ref{Fig1} and Fig. \ref{Fig1_2}, RS-assisted OBF might need a higher $N$ to outperform BS-assisted OBF with the same $M$. This is studied in Fig. \ref{Fig3}, which plots the sum-rate under BS-assisted and RS-assisted OBF using \eqref{AR} under Rician fading channels. The performance achieved using BS-assisted OBF with $M=2$, $4$, and $8$ antennas can be achieved using a single-antenna at the BS with $N=6$, $13$, and $31$ passive reflecting elements at the RS. Note that even under per user path loss, MU diversity gains are still apparent, as shown by the higher curves for $K=128$. 



\begin{figure*}[!t]
\begin{minipage}[b]{0.45\linewidth}
\centering
\tikzset{every picture/.style={scale=.95}, every node/.style={scale=.8}}
%
%
\definecolor{mycolor1}{rgb}{0.00000,0.49804,0.00000}%
\definecolor{mycolor2}{rgb}{0.60000,0.20000,0.00000}%
\definecolor{mycolor3}{rgb}{0.74902,0.00000,0.74902}%
\begin{tikzpicture}

\begin{axis}[%
width=\textwidth,
height=.68\textwidth,
scale only axis,
xmin=2,
xmax=32,
xlabel={$N$},
xmajorgrids,
ymin=0,
ymax=2.7,
ylabel={Sum-rate (bps/Hz)},
ymajorgrids,
legend style={at={(axis cs: 2,2.7)},anchor=north west,legend cell align=left,align=left,draw=white!15!black}
]
\addplot [color=mycolor1,solid,line width=1.0pt,mark=o,mark options={solid}]
  table[row sep=crcr]{%
2	0.366854720442958\\
4	0.363970928998357\\
8	0.35860544598376\\
16	0.358643798558411\\
32	0.364394124473612\\
};
\addlegendentry{\scriptsize OS, $M=1$};

\addplot [color=blue,solid,line width=1.0pt,mark=square,mark options={solid}]
  table[row sep=crcr]{%
2	0.647159933621673\\
4	0.64737414811005\\
8	0.645412193906198\\
16	0.651394970709078\\
32	0.631120407582848\\
};
\addlegendentry{\scriptsize BS-assisted OBF, $M=2$};

\addplot [color=mycolor2,solid,line width=1.0pt,mark=diamond,mark options={solid}]
  table[row sep=crcr]{%
2	1.02661819850949\\
4	1.0247836424134\\
8	1.03492998150808\\
16	1.02068219834576\\
32	1.04528537248829\\
};
\addlegendentry{\scriptsize BS-assisted OBF, $M=4$};

\addplot [color=mycolor3,solid,line width=1.0pt,mark=star,mark options={solid}]
  table[row sep=crcr]{%
2	1.67869480602662\\
4	1.70539395413563\\
8	1.67642463060939\\
16	1.68862977721859\\
32	1.67671742746476\\
};
\addlegendentry{\scriptsize BS-assisted OBF, $M=8$};

\addplot [color=red,solid,line width=1.0pt,mark=triangle,mark options={solid,rotate=270}]
  table[row sep=crcr]{%
2	0.459851955253103\\
4	0.569636238941018\\
8	0.790781570877676\\
16	1.14775340418114\\
32	1.71260536694143\\
};
\addlegendentry{\scriptsize RS-assisted OBF, $M=1$};

\addplot [color=mycolor1,dashed,line width=1.0pt,mark=o,mark options={solid},forget plot]
  table[row sep=crcr]{%
2	0.465295037348723\\
4	0.468181591552571\\
8	0.469967472752067\\
16	0.470122981072972\\
32	0.46627902330123\\
};

\addplot [color=blue,dashed,line width=1.0pt,mark=square,mark options={solid},forget plot]
  table[row sep=crcr]{%
2	0.827172576016182\\
4	0.817963208247329\\
8	0.817230028023793\\
16	0.827238970255615\\
32	0.822441448802849\\
};

\addplot [color=mycolor2,dashed,line width=1.0pt,mark=diamond,mark options={solid},forget plot]
  table[row sep=crcr]{%
2	1.33456834744602\\
4	1.32281447602128\\
8	1.3219654394165\\
16	1.32677471941002\\
32	1.3222489506337\\
};

\addplot [color=mycolor3,dashed,line width=1.0pt,mark=star,mark options={solid},forget plot]
  table[row sep=crcr]{%
2	1.99842622706597\\
4	1.99797970819889\\
8	1.97991231305549\\
16	2.00225746699537\\
32	2.013556852914\\
};

\addplot [color=red,dashed,line width=1.0pt,mark=triangle,mark options={solid,rotate=270},forget plot]
  table[row sep=crcr]{%
2	0.566596919454411\\
4	0.68206670593219\\
8	0.894192271249151\\
16	1.28199183099661\\
32	1.86055910820745\\
};

\node at (axis cs: 16,2.4) [anchor=west] {\scriptsize Solid: $K=64$};
\node at (axis cs: 16,2.2) [anchor=west]{\scriptsize Dashed: $K=128$};
\end{axis}
\end{tikzpicture}%
\caption{Sum-rate in independent Rician channels with per user path loss.}
\label{Fig3}
\end{minipage}
\hspace{.8cm}
\begin{minipage}[b]{0.45\linewidth}
\centering
\tikzset{every picture/.style={scale=.95}, every node/.style={scale=.8}}
%
%
\definecolor{mycolor1}{rgb}{1.00000,0.00000,1.00000}%
\definecolor{mycolor2}{rgb}{0.16471,0.38431,0.27451}%
\begin{tikzpicture}

\begin{axis}[%
width=\textwidth,
height=.68\textwidth,
scale only axis,
xmin=0,
xmax=0.8,
xlabel={$\eta$},
xmajorgrids,
ymin=2.5,
ymax=5.6,
ylabel={Sum-rate (bps/Hz)},
ymajorgrids,
legend style={at={(axis cs: .8,2.5)},anchor=south east,legend cell align=left,align=left,draw=white!15!black}
]

\addplot [color=black,solid,line width=1.0pt,mark size=2pt,mark=triangle,mark options={solid,rotate=270}]
  table[row sep=crcr]{%
0	2.66103028243714\\
0.1	2.66382429475048\\
0.2	2.66485798360953\\
0.3	2.65230642004281\\
0.4	2.6610389777506\\
0.5	2.65607217167395\\
0.6	2.65848725468447\\
0.7	2.65832154784742\\
0.8	2.65824500751366\\
};
\addlegendentry{\scriptsize Rayleigh, OS, $M=1$};

\addplot [color=red,solid,line width=1.0pt,mark size=2pt,mark=star,mark options={solid}]
  table[row sep=crcr]{%
0	4.08272432020589\\
0.1	4.07039254165971\\
0.2	4.08993568809664\\
0.3	4.09317921426317\\
0.4	4.07490818168953\\
0.5	4.06919451845728\\
0.6	4.06979674148668\\
0.7	4.07685360208188\\
0.8	4.08395573888884\\
};
\addlegendentry{\scriptsize Rayleigh, RS-assisted OBF, $M=1$};

\addplot [color=blue,solid,line width=1.0pt,mark size=2pt,mark=o,mark options={solid}]
  table[row sep=crcr]{%
0	4.08295385321839\\
0.1	4.08554786400391\\
0.2	4.06837411780978\\
0.3	4.07298843495631\\
0.4	4.05344580904279\\
0.5	4.04342929226361\\
0.6	4.01032148694593\\
0.7	3.99289911229938\\
0.8	3.97977265496328\\
};
\addlegendentry{\scriptsize Corr. Rayleigh, RS-assisted OBF, $M=1$};

\addplot [color=mycolor1,solid,line width=1.0pt,mark size=2pt,mark=square,mark options={solid}]
  table[row sep=crcr]{%
0	4.08272432020589\\
0.1	4.15734550114647\\
0.2	4.26153502295868\\
0.3	4.34238213257645\\
0.4	4.39756636992931\\
0.5	4.46440294323209\\
0.6	4.52964167127289\\
0.7	4.60166950455775\\
0.8	4.66652966499915\\
};
\addlegendentry{\scriptsize Corr. Rayleigh, RS-assisted det. OBF, $M=1$};

\addplot [color=mycolor2,solid,line width=1.0pt,mark size=2pt,mark=diamond,mark options={solid}]
  table[row sep=crcr]{%
0	3.9594076549597\\
0.1	4.04670904089572\\
0.2	4.12902767008226\\
0.3	4.20690174749982\\
0.4	4.28078672804156\\
0.5	4.35107145665839\\
0.6	4.41809055568595\\
0.7	4.4821340600895\\
0.8	4.54345500323073\\
};
\addlegendentry{\scriptsize Cor. \ref{Cor:RS_OBF_Cor_Rayleigh}};

\addplot [color=blue,dashed,line width=1.0pt,mark size=2pt,mark=o,mark options={solid},forget plot]
  table[row sep=crcr]{%
0	4.47381838053532\\
0.1	4.47392290163992\\
0.2	4.45610060002227\\
0.3	4.45136740013316\\
0.4	4.43155809562522\\
0.5	4.43153401259557\\
0.6	4.40503564975887\\
0.7	4.39058517396808\\
0.8	4.35589043781067\\
};

\addplot [color=mycolor1,dashed,line width=1.0pt,mark size=2pt,mark=square,mark options={solid},forget plot]
  table[row sep=crcr]{%
0	4.47099657509497\\
0.1	4.59923266872534\\
0.2	4.74932115630655\\
0.3	4.89299128654038\\
0.4	5.01871369545354\\
0.5	5.14087682155492\\
0.6	5.27760585296673\\
0.7	5.40731547070869\\
0.8	5.52063109100427\\
};

\addplot [color=mycolor2,dashed,line width=1.0pt,mark size=2pt,mark=diamond,mark options={solid},forget plot]
  table[row sep=crcr]{%
0	4.35107145665839\\
0.1	4.49446232108095\\
0.2	4.63381519692571\\
0.3	4.76942330263915\\
0.4	4.90152876335054\\
0.5	5.03033295877762\\
0.6	5.15600479001086\\
0.7	5.27868732909199\\
0.8	5.3985031885895\\
};

\addplot [color=red,dashed,line width=1.0pt,mark size=2pt,mark=star,mark options={solid}]
  table[row sep=crcr,forget plot]{%
0	4.47099657509497\\
0.1	4.46242308803304\\
0.2	4.47563834090269\\
0.3	4.48121301542804\\
0.4	4.47369335751758\\
0.5	4.46770125219177\\
0.6	4.47294615060514\\
0.7	4.4743376181745\\
0.8	4.4735957020682\\
};

\node at (axis cs: .01,3.5) [anchor = west] {\scriptsize Solid: $N=2$};
\node at (axis cs: .01,3.3) [anchor = west] {\scriptsize  Dashed: $N=3$};

\end{axis}
\end{tikzpicture}%
\caption{Sum-rate in independent Rayleigh and correlated (Corr.) Rayleigh channels.}
\label{Fig4}
\end{minipage}
\end{figure*}

The next result in Fig. \ref{Fig4} studies the effect of correlation on the sum-rate of RS-assisted OBF scheme plotted using the expression in \eqref{AR} for $K=128$. The covariance matrix is assumed to be $[\mathbf{R}]_{i,j}=\eta^{|i-j|}$, $i,j=1,\ldots, N$. As expected with uniformly distributed random phases at the RS, the sum-rate of RS-assisted OBF in correlated Rayleigh fading decreases as $\eta$ increases. However the performance is still much better than that of a system without the RS.

On the other hand, under the deterministic $\mathbf{v}$ in \eqref{design} which exploits the structure of $\bar{\mathbf{R}}$, we obtain $\log_2 (N \lambda_{max}+1)$ improvement in the sum-rate over a system without the RS as shown in Fig. \ref{Fig4} and in accordance with Corollary \ref{Cor:RS_OBF_Cor_Rayleigh}. The simulated values of the sum-rate under deterministic $\mathbf{v}$ scale as the analytical expression in Corollary \ref{Cor:RS_OBF_Cor_Rayleigh}, also plotted in the figure. For $N=2$, $\lambda_{max}$ increases from $1$ to $1.8$ as $\eta$ increases from $0$ to $0.8$. We see a corresponding increase of $\log_2 \frac{1.8N+1}{N+1}=0.62$ in the sum-rate of RS-assisted deterministic OBF over RS-assisted OBF in Rayleigh environment. Therefore by designing $\mathbf{v}$ using the eigenvalue decomposition of $\bar{\mathbf{R}}$, RS-assisted OBF performs better under correlated Rayleigh fading than independent Rayleigh fading. 



\begin{figure*}[!t]
\begin{minipage}[b]{0.45\linewidth}
\centering
\tikzset{every picture/.style={scale=.95}, every node/.style={scale=.8}}
%
%
\definecolor{mycolor1}{rgb}{0.00000,0.49804,0.00000}%
\definecolor{mycolor2}{rgb}{0.85098,0.32549,0.09804}%
\definecolor{mycolor3}{rgb}{1.00000,0.00000,1.00000}%
\begin{tikzpicture}

\begin{axis}[%
width=\textwidth,
height=.68\textwidth,
scale only axis,
xmin=1,
xmax=256,
xmode=log,
xlabel={$K$},
xmajorgrids,
ymin=0,
ymax=19,
ylabel={Sum-rate (bps/Hz)},
ymajorgrids,
legend style={at={(axis cs: 1,19)},anchor=north west,legend cell align=left,align=left,draw=white!15!black}
]
\addplot [color=mycolor1,solid,line width=1.0pt,mark size=2.0pt,mark=o,mark options={solid}]
  table[row sep=crcr]{%
1	0.836801831949919\\
2	1.22445240386659\\
4	1.51158682069552\\
8	1.82908907266148\\
16	2.06088616893101\\
32	2.27897158557747\\
64	2.48708416589784\\
128	2.65846394989036\\
256	2.81767638391828\\
};
\addlegendentry{\footnotesize OS, $M=1$};

\addplot [color=blue,solid,line width=1.0pt,mark size=2pt,mark=diamond,mark options={solid}]
  table[row sep=crcr]{%
1	0\\
2	0.759707388138909\\
4	1.25477201752352\\
8	1.62266874114734\\
16	1.91555482987859\\
32	2.15889793477275\\
64	2.36705875081923\\
128	2.54893723042887\\
256	2.71043230520344\\
};
\addlegendentry{\footnotesize  Ana. result \eqref{TDMA_ray} $\times L$};

\addplot [color=mycolor2,solid,line width=1.0pt,mark size=2.0pt,mark=star,mark options={solid}]
  table[row sep=crcr]{%
1	1.78736616611674\\
2	2.31557299572927\\
4	2.76990999645712\\
8	3.1918483928429\\
16	3.47988023748559\\
32	3.76136283492255\\
64	3.98138377248618\\
128	4.18511789628548\\
256	4.35325731536279\\
};
\addlegendentry{\footnotesize RS-assisted OBF, $N=2, M=1$};

\addplot [color=mycolor3,solid,line width=1.0pt,mark size=2.0pt,mark=x,mark options={solid}]
  table[row sep=crcr]{%
1	0\\
2	1.62266874114734\\
4	2.36705875081923\\
8	2.85565581127947\\
16	3.21998412585876\\
32	3.51060850535784\\
64	3.7523899333058\\
128	3.9594076549597\\
256	4.14041321945906\\
};
\addlegendentry{\footnotesize Cor. \ref{Cor:RS_OBF_Rayleigh_OFDMA}};

\addplot [color=mycolor1,dotted,line width=1.0pt,mark size=2pt,mark=o,mark options={solid},forget plot]
  table[row sep=crcr]{%
1	3.4593036690406\\
2	4.86437073740426\\
4	6.1033327700796\\
8	7.30259136838011\\
16	8.29598738576089\\
32	9.17142273622061\\
64	9.95868413059545\\
128	10.6028888853703\\
256	11.2766949951112\\
};
\addplot [color=blue,dotted,line width=1.0pt,mark size=2pt,mark=diamond,mark options={solid},forget plot]
  table[row sep=crcr]{%
1	0\\
2	3.03882955255563\\
4	5.01908807009409\\
8	6.49067496458937\\
16	7.66221931951436\\
32	8.63559173909102\\
64	9.46823500327691\\
128	10.1957489217155\\
256	10.8417292208138\\
};
\addplot [color=mycolor2,dotted,line width=1.0pt,mark size=2.0pt,mark=star,mark options={solid},forget plot]
  table[row sep=crcr]{%
1	6.63016428448013\\
2	8.94605989062837\\
4	10.9391216630526\\
8	12.4648698410734\\
16	13.631153766293\\
32	14.7714584123423\\
64	15.6325217010538\\
128	16.405277714003\\
256	17.0958552555457\\
};
\addplot [color=mycolor3,dotted,line width=1.0pt,mark size=2.0pt,mark=x,mark options={solid},forget plot]
  table[row sep=crcr]{%
1	0\\
2	6.49067496458937\\
4	9.46823500327691\\
8	11.4226232451179\\
16	12.879936503435\\
32	14.0424340214314\\
64	15.0095597332232\\
128	15.8376306198388\\
256	16.5616528778363\\
};

\node at (axis cs: 1.1,11.5) [anchor = west, fill = white] {\footnotesize Solid: $L=1$};
\node at (axis cs: 1.1,10.3) [anchor = west, fill = white] {\footnotesize Dotted: $L=4$};
\end{axis}
\end{tikzpicture}%
\caption{Sum-rate in a wide-band channel under Rayleigh fading.}
\label{Fig5}
\end{minipage}
\hspace{0.8cm}
\begin{minipage}[b]{0.45\linewidth}
\centering
\tikzset{every picture/.style={scale=.95}, every node/.style={scale=.8}}
%
%
\definecolor{mycolor1}{rgb}{0.74902,0.00000,0.74902}%
\definecolor{mycolor2}{rgb}{0.63529,0.07843,0.18431}%
\definecolor{mycolor3}{rgb}{0.00000,0.49804,0.00000}%
\definecolor{mycolor4}{rgb}{0.87059,0.49020,0.00000}%
\begin{tikzpicture}

\begin{axis}[%
width=\textwidth,
height=.68\textwidth,
scale only axis,
xmin=1,
xmax=256,
xmode = log,
xlabel={$K$},
xmajorgrids,
ymin=0.5,
ymax=5.8,
ylabel={Sum-rate (bps/Hz)},
ymajorgrids,
legend style={at={(axis cs: 1,5.8)},anchor=north west,legend cell align=left,align=left,draw=white!15!black}
]
\addplot [color=red,solid,line width=1.0pt,mark size=2.0pt,mark=o,mark options={solid}]
  table[row sep=crcr]{%
1	0.843274984328077\\
2	1.20189331444475\\
4	1.53157055433485\\
8	1.79304809094217\\
16	2.07561787646302\\
32	2.30090818925701\\
64	2.49718004815283\\
128	2.63616066556011\\
256	2.81742889527724\\
};
\addlegendentry{\footnotesize OS, $M=1$};

\addplot [color=blue,solid,line width=1.0pt,mark size=2pt,mark=square,mark options={solid}]
  table[row sep=crcr]{%
1	1.32295703549308\\
2	1.81522681526892\\
4	2.20759515724633\\
8	2.59427061920742\\
16	2.8956956480616\\
32	3.13554041280017\\
64	3.34593595615782\\
128	3.54772910253688\\
256	3.69762325329485\\
};
\addlegendentry{\footnotesize RS-assisted OBF, $M=1,N=2$};

\addplot [color=mycolor1,solid,line width=1pt,mark size=2.0pt,mark=asterisk,mark options={solid}]
  table[row sep=crcr]{%
1	0.888216607204223\\
2	1.20032774377326\\
4	1.55207251120793\\
8	1.78904871389279\\
16	2.08965789899111\\
32	2.30157545109293\\
64	2.48765892555184\\
128	2.6810013132948\\
256	2.8181543167238\\
};
\addlegendentry{\footnotesize BS-assisted OBF, $M=2$};

\addplot [color=mycolor2,solid,line width=1.0pt,mark size=2.0pt,mark=star,mark options={solid}]
  table[row sep=crcr]{%
1	1.9384365937827\\
2	2.55302340653534\\
4	3.01693337602543\\
8	3.43777750357913\\
16	3.75740335940728\\
32	4.02552565074096\\
64	4.22847028070774\\
128	4.44228364270364\\
256	4.6118834936691\\
};
\addlegendentry{\footnotesize RS-assisted OBF, $M=2, N=2$};

\addplot [color=mycolor3,dashed,line width=1.0pt,mark size=2.0pt,mark=x,mark options={solid}]
  table[row sep=crcr]{%
1	0.518455020860999\\
2	0.724905198971126\\
4	0.982059853919615\\
8	1.20951158438908\\
16	1.38222403959742\\
32	1.56863513220015\\
64	1.73969047351602\\
128	1.87875297034968\\
256	2.00774096242441\\
};
\addlegendentry{\footnotesize BS-assisted OBF, $P=1, M=2$};

\addplot [color=mycolor4,dashed,line width=1.0pt,mark size=2pt,mark=diamond,mark options={solid}]
  table[row sep=crcr]{%
1	1.23569850311242\\
2	1.75497518118972\\
4	2.13462231841641\\
8	2.47733044248411\\
16	2.7466037804935\\
32	3.00213063185012\\
64	3.21932214623018\\
128	3.39539334802924\\
256	3.54292464517556\\
};
\addlegendentry{\footnotesize RS-assisted OBF, $P=1, M=2, N=2$};

\end{axis}
\end{tikzpicture}%
\caption{Sum-rate for multi-user MISO BC under BS-assisted and RS-assisted OBF.}
\label{Fig6}
\end{minipage}
\end{figure*}

We now study the sum-capacity of RS-assisted OBF in a wide-band channel with $L$ sub-carriers under Rayleigh fading. In Fig. \ref{Fig5} we plot both the simulated sum-rate using \eqref{AR} under the channel in \eqref{ch_ov1} for Rayleigh fading $\mathbf{h}_{2,k}^{(l)}$s as well as the asymptotic sum-rate in Corollary \ref{Cor:RS_OBF_Rayleigh_OFDMA}. The simulated values get closer to analytical values as $K$ increases. Moreover, RS-assisted OBF with only $2$ reflecting elements yields a significant improvement over the conventional single-antenna OS system with no RS. Increasing $L$ to $4$ quadruples the sum-rate of both the conventional system and RS-assisted system, since $L$ users are served in each time-slot instead of one.



Finally, we simulate the sum-rate for the MISO BC with $M$ antennas at the BS under BS-assisted OBF as well as RS-assisted OBF for  Rayleigh fading channels. Channel whitening is applied at the BS for the scenario where transmission takes place over the RS-assisted link to remove the spatial correlation as explained in Sec. \ref{Sec:MISO_BS}. The generation of LoS channel matrix $\mathbf{H}_1$ is explained in Sec. \ref{Sec:MISO_BS} while $\mathbf{h}_{2,k}$s undergo independent Rayleigh fading. Introducing an RS into the MISO BC has the positive effect of an improved array gain of $MN$ whereas the negative effect of the introduction of correlation into the channel due to the LoS channel $\mathbf{H}_1$, as highlighted in Theorem \ref{Thm:RS_OBF_Rayleigh_MISOBC}. Fig. \ref{Fig6} shows that the positive effect of the improved array gain outweighs the reduction in sum-rate caused by the introduction of correlation, which is why the MISO system yields a better performance under RS-assisted OBF, and this performance improves as $M$ increases whereas it does not increase under the conventional BS-assisted OBF system. These results are in accordance with the scaling law in Theorem \ref{Thm:RS_OBF_Rayleigh_MISOBC}, which was derived under the assumption that the total available power at the BS scales with $M$. However, if the power constraint in Remark \ref{Rem:Scaling_Power_MISO_BC} is utilized, i.e. total available Tx power $P=1$, the array gain is only $N$ which was already realized by the single-antenna system studied in Theorem \ref{Thm:RS_OBF_Rayleigh}. In fact there is a loss in the performance under RS-assisted OBF due to the introduction of correlation into the channel when multiple antennas are used at the BS. Moreover, there is also a loss in the performance of the conventional MISO system as the average SNR in \eqref{TDMA_ray} will be scaled by $\frac{1}{M}$. Therefore, under the constraint $P=1$, incorporating multiple antennas at the BS of a system that uses OS reduces the sum-rate from what is achieved by the single-antenna system.    

\vspace{-.1in}
\section{Conclusion}

MU diversity has the potential to improve the sum-rate achieved by wireless communication systems by scheduling a user's transmissions at times when its channel SNR is near its peak. In practice, the MU diversity gain is limited in scenarios with strong LoS channel components or spatial correlation. Previous works have proposed OBF using multiple dumb antennas at the BS to enhance the MU diversity gain in these environments. These works focus on the scenario, where a single user is served in each time-slot based on the OS strategy. In this work, we have proposed  OBF using a dumb RS, to induce   large fluctuations in the SNRs of the users by deploying a RS, composed of passive reflecting elements, close to the BS. We show, by deriving the sum-rate scaling laws in the large number of users regime, that this technique amplifies the possible MU diversity gain in Rayleigh, Rician and correlated Rayleigh environments, by a larger factor than BS-assisted OBF while using only a single-antenna at the BS, rendering it to be highly energy efficient. We extended our results to OFDMA systems where multiple users are served on orthogonal frequency bands in each time-slot and MU diversity is exploited in both frequency and time. We also extended the results to MISO BC under OS strategy, and showed that the gain provided by multiple antennas can be easily achieved by using a single antenna at the BS and a higher number of reflecting elements at the RS. Important future research directions include devising fair schedulers for RS-assisted OBF and studying the MU diversity gains provided by RSs in the SDMA system where multiple antennas are used at the BS to serve multiple users in each time-slot using beamforming transmission methods.

\vspace{-.1in}

\appendices

\section{Proof of Theorem \ref{Thm:RS_OBF_Rayleigh} and Corollary \ref{Cor:RS_OBF_Rayleigh}}
\label{App:Proof_RS_OBF_Rayleigh}
Since the users' SNRs $\gamma_{k}$ are i.i.d. so we can study the behaviour of $\gamma_{\hat{k}}=\max_{k\in\{1,\ldots, K\}} \gamma_{k}$ using Lemma \ref{Lem:EVT}. Using \eqref{pdf_ray} in Lemma \ref{Lem:EVT}, we obtain $g(x)=\frac{1-F_{\gamma}(x)}{f_{\gamma}(x)} =\rho_R \beta^2 N >0$, thereby satisfying the condition in \eqref{growth}. Therefore $\gamma_{\hat{k}}-l_{K}$ converges in distribution to a limiting RV $z$. Solving for $l_K$ from Lemma \ref{Lem:EVT}, we obtain $l_{K}=\rho_R \beta^2 N \log K$. Therefore under independent Rayleigh fading, the maximum SNR $\gamma_{\hat{k}}$ grows like $\rho_R \beta^2 N \log K$, which is a function of $N$, $K$, $\beta$ and $\rho$. Using this growth rate of $\gamma_{\hat{k}}$ in \eqref{AR}, we obtain Theorem \ref{Thm:RS_OBF_Rayleigh}.

The proof of Corollary \ref{Cor:RS_OBF_Rayleigh} is similar. We just need to note that under the channel in \eqref{ch_ov}, the pdf and cdf of $\gamma_k$ are given as $f_{\gamma}(x)=\frac{1}{(\rho_R \beta^2 N+\rho_B)} e^{-\frac{x}{(\rho_R \beta^2 N+\rho_B)}}$ and $F_{\gamma}(x)=1-e^{-\frac{x}{(\rho_R \beta^2 N+\rho_B)}}$, respectively. Repeating the same steps as above leads to Corollary \ref{Cor:RS_OBF_Rayleigh}.

\section{Proof of Theorem \ref{Thm:RS_OBF_Rician}}
\label{App:Proof_RS_OBF_Rician}

Recall from \eqref{sum} that the overall channel $h_{k}(t)$ under Rician fading $\mathbf{h}_{2,k}$ is expressed as $h_{R,k}(t)=\sqrt{\rho_R} (\bar{a}_{k}(t)+ c_k(t))$, where $\bar{a}_{k}(t)=\beta \sum_{n=1}^N \sqrt{a}e^{j (\vartheta_{h_1,n}+\theta_{n}(t)+\phi_{k,n})}$, $\vartheta_{h_1,n}=(n-1)2\pi d\sin(\vartheta_n)$ (from the definition of $\mathbf{h}_1$ in Sec. \ref{Sec:Model_and_OBF}), and $c_k(t)=\beta\sum_{n=1}^N h_{1,n}e^{j\theta_{n}(t)} b_{k,n}(t)$. Note that $c_k(t)\sim \mathcal{CN}(0, N\beta^2 u)$ since it is a sum of $N$ independent circularly symmetric complex Gaussian RVs with mean $0$ and variance $\beta^2u$, and that $c_k$ is independent of $\theta_n$s and i.i.d. across $k$. Also note that $|\bar{a}_{k}(t)|\leq N\beta \sqrt{a}$, with equality if the phase allocations at the RS are in beamforming configuration with respect to the fixed component of the channel gain of the user, i.e., $\theta_{n}(t)+\vartheta_{h_1,n}+\phi_{k,n}=\nu(t)\text{ mod} 2\pi$, where $\nu(t)$ is an arbitrary phase. In the following, we drop time-index for simplicity.

Given the phases introduced by the RS have a uniform distribution, then in a system with large $K$, for any fixed $\delta>0$ and $\epsilon \in (0,1)$, and $\forall t$, there exists almost surely a set of $\epsilon K$ users with $\left|\bar{a}_{k}(t)\right| > N\beta \sqrt{a}-\delta$ \cite{dumb}. This happens at every $t$ with a possibly different subset of users. These $\epsilon K$ users can be thought of as experiencing Rician fading with the norm of the fixed component $\approx N\beta \sqrt{a}$. To obtain a lower bound on the sum-rate scaling, we first assume that all the users in this set have the norm of fixed component equal to $N\beta \sqrt{a}-\delta$. The channel gain $|\bar{a}_{k}(t)+ c_k(t)|^2$ of each user in this set will then be distributed as noncentral-$\chi^2(2)$ with non-centrality parameter $(N\beta \sqrt{a}-\delta)^2$ and scaled by $N \beta^2 u/2$. All users' SNRs, $|h_{R,k}|^2$, are therefore i.i.d. and the explicit pdf and cdf are as follows, 
\begin{align}
f_{\gamma}(x)&\hspace{-.5mm}=\hspace{-.5mm}\frac{1}{\rho_R N \beta^2 u} e^{-\frac{(N\beta \sqrt{a}-\delta)^2+\frac{x}{\rho}}{N\beta^2 u}} I_{o}\left(\tau\right), \nonumber \\
F_{\gamma}(x)&\hspace{-.5mm}=\hspace{-.5mm}1\hspace{-.5mm}-\hspace{-.5mm}e^{-\frac{(N\beta \sqrt{a}-\delta)^2+\frac{x}{\rho_R}}{N\beta^2 u}} \hspace{-.5mm}\sum_{m=0}^{\infty} \hspace{-.9mm}\frac{\left(N\beta \sqrt{a}\hspace{-.5mm}-\hspace{-.5mm}\delta\right)^m}{(x/\rho_R)^{\frac{m}{2}}}  I_{m}(\tau), \nonumber
\end{align}
for $x>0$, where $\tau=\frac{\sqrt{\frac{x}{\rho_R}}(N\beta \sqrt{a}-\delta)}{N\beta^2 u/2}$.


The proof then follows from approximating the tails of $f_{\gamma}(x)$ and $F_{\gamma}(x)$ by using  $I_{k}(x)\sim e^{x}/\sqrt{2\pi x}$ and using $\sum_{m=0}^{\infty} \frac{(N\beta \sqrt{a}-\delta)^m }{(x/\rho_R)^{\frac{m}{2}}}\to1$ 
as $x\to\infty$. This yields
\begin{align}
\label{pdf_ric_app}
&\hspace{-1.5mm}f_{\gamma}(x) \hspace{-.5mm}\sim\hspace{-.5mm} \frac{e^{-\left(\sqrt{\frac{x}{\rho_R}}-(N\beta \sqrt{a}-\delta)\right)^2/N\beta^2 u}}{2\sqrt{\pi N\beta^2 u(N\beta\sqrt{a}-\delta)\sqrt{x\rho_R^3}} } , \\
\label{step}
&\hspace{-1.5mm}F_{\gamma}(x) \hspace{-.5mm}\sim \hspace{-.5mm}1 \hspace{-.5mm}-\hspace{-.5mm} \frac{\sqrt{\hspace{-.5mm}N\beta^2 u}e^{-\left(\hspace{-.5mm}\sqrt{\frac{x}{\rho_R}}-(N\beta \sqrt{a}-\delta)\right)^{\hspace{-.5mm}2}\hspace{-1.0mm}/N\beta^2 u}}{ 2\sqrt{\pi(N\beta \sqrt{a}-\delta)\sqrt{x/\rho_R}}},
\end{align} 
as $x\to\infty$. We can study the lower bound on the growth rate of the maximum of the SNRs among these $\epsilon K$ users for large $K$ using Lemma \ref{Lem:EVT}. Inserting \eqref{pdf_ric_app} and \eqref{step} into \eqref{growth}, we obtain $\underset{x\to\infty}{\text{lim}}\frac{1-F_{\gamma}(x)}{f_{\gamma}(x)} =\rho_R N \beta^2 u >0$, thereby satisfying the condition in \eqref{growth}. Therefore $\gamma_{\hat{k}}-l_{\epsilon K}$ converges in distribution to a limiting RV $z$. Solving for $l_{\epsilon K}$ from Lemma \ref{Lem:EVT}, we end up obtaining $\sqrt{\frac{l_{\epsilon K}}{\rho_R}}=\sqrt{\mu}+(N\beta \sqrt{a}-\delta)$ where $\mu=N\beta^2u(\log  K +\log \epsilon) - N\beta^2u \log \Big(2\sqrt{\frac{\pi}{ N \beta^2 u }}\Big( \frac{l_{\epsilon K}}{\rho_R}(N\beta \sqrt{a} \\ -\delta)^2 \Big)^{1/4}\Big)$.
The maximum of the SNRs among these $\epsilon K$ users grows  at least as fast as $l_{\epsilon K}= \rho_R(\sqrt{N\beta^2 u\log K}+(N \beta \sqrt{a}-\delta))^2+O(\log \log K)$ as $K\to\infty$, for fixed $\epsilon,\delta>0$. Since this is true for any $\delta>0$ and for a subset of users, we conclude that a lower bound on the growth rate of $\gamma_{\hat{k}}$ is $l_K=\rho_R N\beta^2(\sqrt{u\log K}+\sqrt{Na})^2+O(\log \log K)$. This growth rate is also attained by the ideal situation where all users are simultaneously at the beamforming configurations of their fixed component, i.e. have the norm of fixed component $\left|\bar{a}_{k}(t)\right|$ equal to $N\beta \sqrt{a}$. Using this interpretation, $l_K$ can also be shown to be an upper bound to the growth rate of $\gamma_{\hat{k}}$. Thus, the scaling of sum-capacity under RS-assisted OBF is obtained by substituting $l_K=\rho_RN\beta^2 (\sqrt{u\log K}+\sqrt{Na})^2+O(\log \log K)$ for $\underset{k=1,\ldots,K}{\max} \gamma_{k}$ in \eqref{AR}. This proves Theorem \ref{Thm:RS_OBF_Rician}.

\section{Proof of Lemma \ref{Lem:Cor_Rayleigh}}
\label{Proof_Cor_Rayleigh}
We first provide the cdf of $\gamma_{k}$ given $\mathbf{v}$. Since all users' SNRs are i.i.d., we will drop the index $k$. Given $\mathbf{v}$, the cdf is expressed as $F_{\gamma}(x)=P[\gamma \leq x]=P[\rho_R \mathbf{z}^H \mathbf{A}  \mathbf{z} \leq x]=P[\rho_R \mathbf{z}^H \mathbf{Q}^H \boldsymbol{\Lambda}\mathbf{Q}  \mathbf{z} \leq x]$, where $\mathbf{A}=\mathbf{Q}\boldsymbol{\Lambda} \mathbf{Q}^H$ is the eigenvalue decomposition of $\mathbf{A}$. Note that $\mathbf{A}$ is a rank-one matrix so it will only have one non-zero eigenvalue $\lambda=\bar{\mathbf{v}}^H \bar{\mathbf{R}} \bar{\mathbf{v}}$. Denoting $\bar{\mathbf{z}}=\mathbf{Q}\mathbf{z}$, we express the cdf as $F_{\gamma}(x)=P[\rho_R \bar{\mathbf{z}}^H  \boldsymbol{\Lambda} \bar{\mathbf{z}} \leq x]=P[\rho_R \bar{z}^* \lambda \bar{z} \leq x]$,
where $\bar{z}$ denotes the component of $\bar{\mathbf{z}}$ corresponding to the only non-zero component of $\boldsymbol{\Lambda}$, which is $\lambda$. Note that the components of $\bar{\mathbf{z}}$ are i.i.d. $\mathcal{CN}(0,1)$ so $\rho_R \lambda  \bar{z}^* \bar{z}$ is an exponential RV with parameter $\frac{1}{\rho_R \lambda}$. The closed-form expressions of the cdf and pdf of $\gamma$ given $\mathbf{v}$ are thus $F_{\gamma}(x)= 1-e^{-\frac{x}{\rho_R \lambda}}$ and $f_{\gamma}(x)=\frac{1}{\lambda \rho_R} e^{-\frac{x}{\lambda \rho_R}}$, with $x\geq 0$, respectively.

Next we develop the scaling law of maximum SNR, $\gamma_{\hat{k}}=\underset{k=1,\ldots, K}{\max} \gamma_{k}$, as $K$ grows large using Lemma \ref{Lem:EVT}. Inserting the pdf and cdf into Lemma \ref{Lem:EVT}, we obtain $g(x)=\frac{1-F_{\gamma}(x)}{f_{\gamma}(x)} =\rho_R \lambda >0$, thereby satisfying \eqref{growth}. Therefore $\gamma_{\hat{k}}-l_{K}$ converges in distribution to a limiting RV $z$. Solving for $l_K$ from Lemma \ref{Lem:EVT}, we have $e^{-\frac{l_K}{\rho_R \lambda}}=\frac{1}{K}$ and therefore $l_{K}=\rho_R \lambda \log K$. Substituting $l_K$ for $\underset{k=1,\ldots, K}{\max} \gamma_{k}$ in \eqref{AR2}, we obtain Lemma \ref{Lem:Cor_Rayleigh}.

\bibliographystyle{IEEEtran}
\bibliography{bib}
\end{document}